\documentclass[letterpaper, 10 pt, onecolumn]{ieeeconf} 
\usepackage[left=3.5truecm,right=3.5truecm,top=3.25truecm,bottom=3truecm]{geometry}
\IEEEoverridecommandlockouts

\usepackage{amsfonts, dsfont, mathtools, amssymb, bbold, mathrsfs, amsmath, verbatim}
\usepackage{graphicx, float, epsfig, color, psfrag, adjustbox, wrapfig}
\usepackage[table]{xcolor}
\usepackage[noend]{algpseudocode}
\usepackage[ruled,vlined]{algorithm2e}
\usepackage{subcaption}
\usepackage{refcount, url}
\usepackage{hyperref, cleveref}

\def\R{{\mathds{R}}}

\def\0{{\mathbb{0}}}
\def\1{{\mathds{1}}}

\def\H{{\mathcal H}}   

\def\r{{\mathbf{r}}}
\def\v{{\mathbf{v}}}

\SetKwInOut{Parameter}{Parameters}

\newcommand{\norm}[1]{\left\lVert#1\right\rVert}

\newtheorem{theorem}{\bf Theorem}[section]

\newtheorem{remark}{Remark}[section]

\graphicspath{{figures/}, {figures_entropy}}

\title{\LARGE \bf
Flocks, Games, and Cognition: A Geometric Approach}

\author{
Udit Halder$^{1, \dagger}$, Vidya Raju$^{2, \dagger}$, Matteo Mischiati$^{3}$, Biswadip Dey$^{4}$, P.~S.~Krishnaprasad$^{5,6, \star}$
\thanks{This research was supported by the Air Force Office of Scientific Research under AFOSR Grant FA9550-10-1-0250, an AFOSR FY2012 DURIP Grant No. FA2386-12-1-3002, the ARL/ARO MURI Program Grant No. W911NF-13-1-0390, through the University of California Davis (as prime), the ARL/ARO Grant No. W911NF-17-1-0156, through the Virginia Polytechnic Institute and State University (as prime), the National Science Foundation CPS: Medium Grant CNS-2013824 through University of California Irvine (as prime), and by Northrop Grumman Corporation.}
\thanks{$^{1}$Coordinated Science Laboratory, University of Illinois, Urbana-Champaign, IL 61801, USA, 
$^2$John A. Paulson School of Engineering and Applied Sciences, Harvard University, MA 02138, USA, 
$^{3}$Independent Researcher, Ashburn, VA 20147, USA, 
$^{4}$Siemens Corporation, Technology, Princeton, NJ 08540, USA, 
$^{5}$Institute for Systems Research and $^{6}$Department of Electrical and Computer Engineering, University of Maryland, College Park, MD 20742, USA.}
\thanks{$^\dagger$Equal contributions}
\thanks{$^\star$Corresponding e-mail:  {\tt\small krishna@umd.edu}}
}

\begin{document}
\maketitle

\begin{abstract}
Avian flocks display a wide variety of flight behaviors, including steady directed translation of center of mass, rapid change of overall morphology, re-shuffling of positions of individuals within a persistent form, etc. These behaviors may be viewed as flock-scale \textit{strategies}, emerging from interactions between individuals, accomplishing some collective \textit{adaptive} purpose such as finding a roost, or mitigating the danger from predator attacks. While we do not conceive the flock as a single cognitive agent, the moment-to-moment decisions of individuals, influenced by their neighbors, appear \textit{as if} to realize collective strategies that are cognizant of purpose. In this paper, we identify the actions of the flock as allocation of energetic resources, and thereby associate a \textit{cognitive cost} to behavior. Our notion of cognitive cost reflects the burden arising from rapid re-allocation of resource. Using a recently developed natural geometric approach to kinetic energy allocation, we map the flock behavior to a temporal signature on the standard (probability) simplex. Given the signature of a flocking event, we calculate the cognitive cost as a solution to an optimal control problem based on a game-theoretic model. Alternatively, one can associate to a signature an \textit{entropic cost}. These two cost measures, when applied to data on starling flocks, show a consistent spread in value across events, and we suggest the possibility that higher cost may arise from predator attacks. 
\end{abstract}


\section{Introduction}
In modern studies of avian flocks, significant progress has been made, thanks to advances in motion capture technologies based on computer vision, as in the work of Cavagna and collaborators \cite{Ballerini_TOPO, ballerini2008empirical,  attanasi2014information, cavagna2018propagating}, and GPS data-logging methods as in the work of Vicsek and collaborators \cite{vicsek2010hierarchical}. While the investigations of starling flocks in \cite{Ballerini_TOPO, ballerini2008empirical} were focused on the structure and statistics of interactions ruling the observed flocking events (a bottom-up approach), questions of flock-scale phenomena, such as information transfer through waves were the subject of \cite{attanasi2014information, cavagna2018propagating,halder2019optimality}. In \cite{vicsek2010hierarchical} the authors were concerned with pigeon flocks engaged in free flight or homing behaviors and the appearance of leadership structures in these settings. In this paper, we are concerned with flock-scale phenomena found in the starling flight data obtained by the Collective Behavior in Biological Systems (COBBS) group, led by Dr. Andrea Cavagna of the Institute for Complex Systems (ISC-CNR) in Rome. Specifically, we use a \textit{top-down} approach to the dynamics of observed flight behaviors, including steady directed translation of center of mass of the flock, rapid change of overall morphology, re-shuffling of positions of individuals within a persistent form, etc. It has been suggested that such behaviors confer anti-predatory advantage on flocks thanks to the \textit{confusion effect} \cite{krakauer1995groups, beauchamp2013social, ruxton2007confusion} -- predators such as peregrine falcons, that are exceptionally successful with isolated targets, are foiled by the perceptual challenges posed by large flocks \cite{hunt2013darwinian}. The time course of fractional allocation of kinetic energy resource among different behaviors can be represented as a signature of a flocking event on a standard simplex. Treating distinct flight behaviors as competing \textit{pure strategies in a game}, the signature is a trace of mixed strategies. We draw on the subject of evolutionary game theory to establish natural families of dynamics and control systems on the simplex. Fitting a generative model from such a family to each signature is formulated as an optimal control problem on the simplex. Solutions to such optimal control problems allow us to propose and examine a notion of cognitive cost of flock behavior.

       Computer vision algorithms yield data on starling flocks in the form of streams of three dimensional coordinates of individuals in a flocking event approximately every 6ms \cite{attanasi2014information, cavagna2014bird}. To extract dynamical properties of trajectories, including velocity and acceleration, it is necessary to smooth the data. While there is a long history of smoothing techniques in biological data analysis, we used an efficient method based on the theory of optimal control of linear systems with quadratic cost functionals, to obtain smoothed trajectories for each bird \cite{dey2012trajectory, dey2015reconstruction}. The passage from individual-scale to flock-scale analysis is based on a recent development of the idea of kinematic modes in many-particle systems \cite{mischiati2017geometric}. Using the geometric language of fiber bundles, the velocity of the flock as a whole is split into several mutually orthogonal components (kinematic modes). The notion of orthogonality is based on a Riemannian metric tensor defined by the masses of individuals (here assumed to be equal on the basis of homogeneity of the flock). We note that our kinematic modes are akin to the chemist's idea of normal modes of vibration of polyatomic molecules in spectroscopy \cite{wilson1980molecular, weinstein1973normal}. The kinetic energy of a flock is in turn split into energy modes (e.g. kinetic energy relative to center of mass, and its further splitting into kinetic energy of flock shape deformation relative to center of mass, kinetic energy of global rigid rotation etc.). Taking fractions of the different energy modes with respect to the kinetic energy relative to center of mass yields a signature of a flocking event on a standard (probability) simplex. In \cite{mischiati2017geometric} this process was applied to pigeon flock data from \cite{vicsek2010hierarchical}.

      The simplex is the natural space for representing mixed strategies in a game with finitely many pure strategies \cite{nash1951non}. In a typical flocking event one does not see sustained pure strategies e.g. global rigid rotation. Instead one finds non-zero fractions of multiple energy modes, which we interpret as the occurrence of mixed strategies. In the development of evolutionary game theory in biology \cite{smith1973logic}, a central idea is that of replicator dynamics \cite{taylor1978evolutionary, smith1982evolution} specified by a vector-valued fitness map on the simplex. It captures the evolution of mixed strategies. In a recent development \cite{raju2019cognitive, raju2020lie}, it has been suggested that such dynamics may be viewed as occupying the middle layer of a three layer cognitive hierarchy where the top (cognitive) layer concerns the control of replicator dynamics through modulation of fitness maps. This is of particular interest, since the observed signature of a flocking event cannot in general be identified with a fixed replicator dynamics due to self-intersections, but one can fit a controlled replicator dynamics to the signature. The control enables time-dependence of energy mode allocation. This leads to a natural optimal control problem on the simplex, to be solved by construction of a Hamiltonian system via the Maximum Principle of Pontryagin and coworkers \cite{pontryagin1962mathematical}. The time-averaged Hamiltonian arising from the optimal solution directly reflects the \textit{rate of re-allocation} of energy resource among different energy modes. For this reason, we interpret the time-averaged Hamiltonian as cognitive cost of flocking. On the other hand one can associate an entropy value to each point on a signature curve, and hence the average entropy of the signature. This reflects the degree of unpredictability of flock behavior and the extent of confusion effect that a predator targeting a flock is subject to. This measure is insensitive to the temporal variation of the flock signature. This paper examines both these measures on starling flock data. In what follows, we discuss the organization and contributions of this paper.

       In Section \ref{sec:data}, we briefly indicate the characteristics of the starling flock data used in this paper. Section \ref{sec:smoothing_linear} is an account of the linear-quadratic optimal control methods \cite{dey2012trajectory, dey2015reconstruction} used to smooth the sampled data to obtain trajectories of individual birds in the flocking events.  In Section \ref{sec:smoothing_euclidean}, a brief summary of a more general approach based on the Pontryagin Maximum Principle (PMP) is given for use later in Section \ref{sec:generative_optimal} in connection with the optimal energy mode allocation problem. In Section \ref{sec:energymodes} the different energy mode splittings originating in \cite{mischiati2017geometric} are described. Section \ref{sec:generative} provides an outline of the concept of cognitive hierarchy developed in \cite{raju2019cognitive} and specializes it in Section \ref{sec:generative_special2} to the setting of games with two pure strategies. This leads to the optimal control problem in the cognitive layer discussed in Section \ref{sec:generative_optimal}. The resulting evaluations of cognitive cost and entropic cost under energy splittings for eight flocking events are presented and compared in Section \ref{sec:results}. The value of such measures lies in ordering or distinguishing flocking events, with the possibility of discerning/suggesting underlying causes for the observed differences. We conclude with discussions and interpretations of these results in Section \ref{sec:discussion}. 

\medskip 
\noindent 
\textbf{Historical Remark}: The cognitive cost of human decision-making has been of interest to researchers in behavioral psychology and neuroscience for quite some time (see \cite{christie2015cognitive} and references therein, as well as \cite{egner2017wiley}). In \cite{christie2015cognitive}, a combination of physiologically plausible models of glycogen metabolism in the brain, an associated optimal control formulation, and simulation experiments are used to put forward a case for energy utilization as a means to account for cognitive cost. While this presupposes that the brain has an intelligent control system to manage the use of its metabolic resources, in the present work we do not wish to suggest any sort of central authority controlling resource allocation in the flock. Indeed one expects that the energy resource (mode) allocation, seen in starling data is an \textit{epiphenomenon} of perceptually-guided steering behaviors of individuals responding to conspecific neighbors and predators (such as peregrine falcons). Individual behaviors may have been selected through evolution thanks to the benefits of predator avoidance conferred by flock-scale behaviors (leading to confusion effect). While the notion of energy resource allocation in our work parallels that in \cite{christie2015cognitive}, we are not dependent on any physiologically based models. Instead our measure of cognitive cost is centered on: (i) it is possible to recognize a finite set of flock-scale behaviors; (ii) there are associated kinetic energy modes; and (iii) rapid resource reallocation across energy modes (modeled via \textit{controlled replicator dynamics}) incurs a burden on the flock (measured as cognitive cost using optimal control theory). In the investigation of signatures of decision-making processes, cognitive scientists have identified a spectrum ranging from the automatic, fast, and relatively inflexible extreme to the slower, flexible, and deliberative extreme, the latter often referred to as ``cognitive control" – a term we do not use in this paper to avoid conflating with control-theoretic usage \cite{rand2017cyclical}. It is important to note that these authors initiate and explore a modeling approach based on replicator dynamics on the 1-dimensional simplex, incorporating various feedback laws, and costs. The fast and slow extremes are treated as pure strategies in a game, akin to the notion of energy modes (with mechanical origins) in our own work. As we show later, the geometric approach of the present work allows consideration of signatures in higher dimensional simplices. Another relevant antecedent for our control-theoretic notion of cognitive cost has to do with the problem of selective attention to sensory input and motor sequences -- of great interest to neuroscience and cognitive science. In pioneering work Roger Brockett formulated a notion of \textit{cost of attention}, which leads to novel problems of infinite dimensional control – specifically (optimal) control on the diffeomorphism group and Liouville equations \cite{brockett2012notes}.

Whereas this paper is aimed at quantitative exploration of flocking events and their dynamic morphologies by considering underlying adaptive purpose (e.g. predation avoidance), the recorded history of observations by naturalists have for long provided questions and suggested qualitative explanations that have stimulated research. In her article \cite{nice1935edmund}, referring to the naturalist Edmund Selous as a confirmed Darwinian, Margaret Nice quotes from his famous book \cite{selous1905bird} this memorable description (on page 141) of a starling flock:

\medskip
{\footnotesize
``...and now, more and faster than the eye can take it in, band grows upon band, the air is heavy with the ceaseless sweep of pinions, till, glinting and gleaming, their weary wayfaring turned to swiftest arrows of triumphant flight—toil become ecstasy, prose an epic song—with rush and roar of wings, with a mighty commotion, all sweep, together, into one enormous cloud. And still they circle; now dense like a polished roof, now disseminated like the meshes of some vast all-heaven-sweeping net, now darkening, now flashing out a million rays of light, wheeling, rending, tearing, darting, crossing, and piercing one another —a madness in the sky."
}
\medskip

In this passage one senses the fascination that the dynamic flock morphology and its likely purpose hold for casual and scholarly observers alike. Further on Selous speculates about the collective guidance that might drive the phenomenon (pages 142-143 in \cite{selous1905bird}). A modern view treats the phenomenon as the result of co-evolution of predator (peregrine falcon) and prey (common starling) -- to quote naturalist Grainger Hunt \cite{hunt2013darwinian}:

\medskip
{\footnotesize
``What are we to make of the pulsating, other-worldly spectacle of a massive starling flock, moving amoeba-like across the open skies? A Peregrine Falcon or other winged predator is almost always involved, as the thousands of individual flock members fight to evade capture."
}
\smallskip

And again,

\medskip
{\footnotesize
``The wondrous cloud [of starlings] is thus secondary – an extraneous property, emerging from independent attempts by each individual, within the multitude of self-interested starlings, to escape the falcon." \cite{hunt2013darwinian} 
}
\medskip

Hunt also views as a relevant factor the self-interest of the falcon in avoiding injury that might result from even a grazing collision with a starling in a flock – thereby raising the failure rate for prey-capture by the falcon.

\medskip
{\footnotesize
``And so the peregrine attacks the flock gingerly, and in apparent moderation of its true ability to catch [an isolated] starling." \cite{hunt2013darwinian}  
}
\medskip

\medskip 
\noindent 
\textbf{Dedication}: We dedicate this paper to Professor Arthur Krener on his eightieth birthday, celebrating his many contributions to nonlinear and optimal control, geometric perspectives, and computational investigation.

\section{Flocking Data}\label{sec:data}
This project grew out of a collaboration between the University of Maryland and the COBBS group of the Institute for Complex Systems in Rome (ISC-CNR), located at the University of Rome ``La Sapienza”. Andrea Cavagna had already pursued an earlier campaign of observations of starling flocks (with support from the European Union, StarFlag 2007), and the Maryland-Rome collaboration was supported by the U.S. Air Force Office of Scientific Research (AFOSR).

Starlings gather around urban areas during winter months to gain extra warmth from the cities. They spend the day feeding in the countryside, and before settling on the trees for the night they gather in flocks to perform elaborate aerial displays. Data on flocking events were captured by the COBBS group during 2011. Three high speed cameras (IDT M5) were used for this purpose, with a maximum frame rate of 170 frames per second (fps) at a resolution of 2288x1728. These time-sampled data were taken from the roof of Palazzo Massimo, Museo Nazionale Romano, in the city center of Rome, in front of one of the major roosting sites used by starlings. Data pertaining to eight flocking events were supplied to the Maryland group (authors of this paper). The details for the particular flocking events used in this paper are to be found in Table \ref{tbl:flock}. See also \cite{attanasi2014information} for additional details pertaining to the dataset.

The data supplied to the Maryland group contained time-stamped 3D coordinates of each bird in each flocking event, obtained using advanced computer vision algorithms developed by the COBBS group. One of the key steps in our analysis is to develop from this data, requisite trajectory characteristics at finer resolution through a process of smoothing discussed in the next section. In particular, velocity, acceleration and jerk were extracted in this way.

The flocking data came unlabeled in that none of the eight events was identified as occasioned by a predator attack even though starling flocks invariably elicit attention from peregrine falcons \cite{hunt2013darwinian}. Our analysis below suggests candidate events that may be so labeled.

\begin{table}[t]
\footnotesize
\centering
  \begin{tabular}{ c | c | c | c }
  \hline
    \rowcolor{black}
    {\color{white} Flocking}  & {\color{white}Flock Size} & {\color{white} Duration} 	& {\color{white}Data Capture Rate} \\
    \rowcolor{black}
    {\color{white}Event}     &   {\color{white}$(n)$}		   & {\color{white}(seconds)}  & {\color{white}(frames/second)}  \\    
    &&&\\[-0.5em]
   1 & 175 & 5.4875 & 80 \\     [0.4em]
   2 & 123 & 1.8176 & 170 \\    [0.4em]
   3 &   46 & 5.6118 & 170 \\ [0.4em]
   4 & 485 & 2.3471 & 170 \\ [0.4em]
   5 & 104 & 3.8824 & 170 \\ [0.4em]
   6 & 122 & 4.1588 & 170 \\ [0.4em]
   7 & 380 & 5.7353 & 170 \\ [0.4em]
   8 & 194 & 1.7588 & 170\\  [0.2em]\hline
  \end{tabular}
  \caption{Details of captured flocking events}
  \label{tbl:flock}
\end{table}

\section{Data Smoothing} \label{sec:smoothing}

Given a time-indexed sequence of sampled observations on a manifold, generative models provide a meaningful way of capturing them through the use of an underlying dynamical system complete with control inputs having useful interpretations. The control inputs are determined by solving an optimal control problem, where the cost function consists of a data-fitting term that penalizes mismatch between the generated trajectory and sampled data, and a smoothing term weighted by a parameter $\lambda$ that affects the smoothness of the generated trajectory. We discuss two generative models to solve this problem. 
\subsection{A Linear Generative Model}
\label{sec:smoothing_linear}
A first approach to solving the data smoothing problem (in 3 dimensions), presented in \cite{dey2012trajectory}, is to formulate an optimal control problem to minimize the jerk path integral, with intermediary state costs determining the fit error. Suppose that $\left\{\textbf{r}_i\right\}_{i=0}^{N}$ denote the positions $\mathbf{r}_i \in \R^3$ of the birds at each sampling time instant $t_i$. In order to recover a trajectory fit $\textbf{r}(t): [t_0,t_N] \rightarrow \R^3$, one can use the jerk-driven linear generative model
\begin{align}
\begin{split}
\dot{\textbf{r}}(t) &= \textbf{v}(t)\\
\dot{\textbf{v}}(t) &= \textbf{a}(t) \\
\dot{\textbf{a}}(t) &= \textbf{u}(t), 
\end{split}
\label{eq:jerk_dynamics}
\end{align} 
where $\textbf{v}(t), \textbf{a}(t), \textbf{u}(t)$ denote the velocity, acceleration, and jerk (input) of the trajectory respectively. The cost functional to be minimized is
\begin{align}
J_l &=\sum\limits_{i=0}^{N} ||\textbf{r}(t_i) - \textbf{r}(t)||^2 + \lambda \int\limits_{t_0}^{t_N}||\textbf{u}(t)||^2\,dt, 
\end{align} 
where $||\cdot||$ denotes the Euclidean norm, and the minimization is over initial conditions $\textbf{r}(t_0),\textbf{v}(t_0),\textbf{a}(t_0)$ and the input function $\textbf{u}(\cdot)$. Defining the state as
\begin{align}
\textbf{x}(t) = \left[\begin{array}{c}
\textbf{r}(t) \\
\textbf{v}(t) \\
\textbf{a}(t) 
\end{array}\right] \in \R^9, 
\end{align}
we can write \eqref{eq:jerk_dynamics} as the linear control system
\begin{align}
\dot{\textbf{x}}(t) &= A\textbf{x}(t) + B\textbf{u}(t),
\label{lingen}
\end{align}
with  the matrices $A$ and $B$ defined appropriately. The problem of minimizing $J_l$ subject to (\ref{lingen}) becomes a linear, quadratic optimal control problem, which can be solved by a completion of squares of terms in the cost by invoking a path independence lemma, or by applying the Pontryagin Maximum Principle as shown in \cite{dey2012trajectory}. This approach has been used to smooth the starling flock data \cite{dey2015reconstruction} for all the events listed in Table~\ref{tbl:flock}, with the parameter $\lambda$ found by leave-one-out cross validation.

%

\subsection{Data Smoothing in the Euclidean Setting}
\label{sec:smoothing_euclidean}
In this section, we present a general result on the Pontyagin Maximum Principle based approach for data smoothing on the Euclidean space $\R^k$. Suppose that $\left\{x_i^d\right\}_{i=0}^{N}$ denote the sampled data. For a generative model given by the dynamics $\dot{x} = f(x,u)$ on ${\R}^k$, with the control $u \in \R^m$, the optimal control problem can be formulated as
\begin{align}
\begin{split}
& \min\limits_{x(t_0),~ u(\cdot)} J(x(t_0), u) = \sum\limits_{i=0}^N F_i (x(t_i),x_i^d) + \frac{\lambda}{2} \int_{t_0}^{t_N} \norm{u(t)}^2 dt,  \\ 
& \text{subject to:~~} \dot{x} = f(x,u), 
\end{split}
\label{eq:optimal_control_problem_gen}
\end{align}
where parameter $\lambda > 0$ is a regularization parameter, and $F_i$'s are suitably defined \textit{fit errors} of the reconstructed trajectories and sampled data at the sampling times. Using Pontryagin Maximum Principle, the optimal control values can be calculated as a function of the state and a co-state variable. The following result from \cite{dey2014control} states this precisely.
\begin{theorem} (PMP for data smoothing \cite{dey2014control} ) Let $u^*(\cdot)$ be an optimal control input for \eqref{eq:optimal_control_problem_gen}, and let $x^*(\cdot)$ denote the corresponding state trajectory. Then there exists a costate trajectory $p : [t_0, t_N] \rightarrow \R^k,\, p \neq 0$, such that 
\begin{align}
\begin{split}
\dot{x}^* &=  \frac{\partial \H}{\partial p} (t, x^*, p, u^*) \\
\dot{p} &= -\frac{\partial \H}{\partial x} (t, x^*, p, u^*), \\
\end{split}
\label{eq:xp_dot_thm}
\end{align}
during $t \in (t_i, t_{i+1}), ~i = 0, 1, ..., N-1$, and the Hamiltonian is given as
\begin{align}
\H(t, x^*, p, u^*) = \max\limits_{v\in \R^m} H(t, x^*, p , v), 
\label{eq:hamiltonian_max}
\end{align}
for $t \in [t_0, t_N] \setminus \{t_0, t_1, ..., t_N\}$, where the pre-Hamiltonian is defined as~$H(t, x, p, u) = p{^\mathsf{T}}f(x,u) - \frac{\lambda}{2} \norm{u}^2$. Moreover, jump discontinuities of the costate variable can be written as
\begin{align}
\begin{split}
& p(t_0^-) = 0, \\
& p(t_i^+) - p(t_i^-) = \frac{\partial F_i (x(t_i))}{\partial x(t_i)}, \quad i = 0, 1, ..., N, \\
& p(t_N^+) = 0.
\end{split}
\label{eq:jump_p_thm}
\end{align} 
\label{thm:PMP_dataSmoothing}
\end{theorem}
The piecewise continuous nature of the co-state trajectory due to jump conditions arising from mismatch between the sampled data points and the reconstructed state must be noted here. The initial condition $x(t_0)$ is identified by using the terminal condition for the co-state, while the optimal value of $\lambda$ is typically obtained through leave-one-out cross validation. The reconstructed trajectory is then obtained as the projection onto the state space of the solution of Hamilton's equations derived from the (maximized pre-) Hamiltonian. We refer the reader to \cite{dey2015reconstruction} for a detailed treatment of this problem. This is the result that will be used in our data fitting problem on a simplex (Section~\ref{sec:generative_optimal}).


\section{Energy Modes} \label{sec:energymodes}
Avian flocks display a variety of flight behaviors that may be characterized as collective strategies such as steady translation of center of mass (which we denote by \textit{com}), coherent rotation about center of mass (\textit{rot}), change of ensemble form (\textit{ens}), internal re-shuffling of relative positions or democratic strategy (\textit{dem}), (rapid) expansion or contraction of volume (\textit{vol}) etc. A flocking event may display all of the mentioned strategies to varying degrees as governed by the time-dependent allocation of kinetic energy to each strategy. We take the viewpoint presented in \cite{mischiati2017geometric} and study the fractions of the total kinetic energy of a flock allocated to several `kinematic modes' -- rigid translations, rigid rotations, inertia tensor transformations, expansion and compression, in order to describe collective behavior. By doing so, we treat the flock as a single entity with several strategies of energy allocations emerging from individual behavior. Below is a brief discussion on the resolution of kinetic energy into components, from \cite{mischiati2017geometric}. 

If the positions of the birds in a flock are denoted by $\{\r_1, \r_2, ..., \r_n \}$, the center of mass can be written as
\begin{align}
\r_{\textbf{com}} = \frac{1}{n}\sum\limits_{i=1}^n \r_i,
\end{align}
where we treat every bird alike, i.e. their masses are taken to be equal to 1. The \textit{ensemble inertia tensor} is defined by 
\begin{align}
K = \sum\limits_{i=1}^n \left(\r_i - \r_{\textbf{com}}\right) \left( \r_i - \r_{\textbf{com}}\right)^{\mathsf{T}}.
\label{eq:K}
\end{align}
Let the velocities of the birds be denoted as, $\{ \v_{\r 1}, ..., \v_{\r n} \}$, then the total kinetic energy is
\begin{align}
E = \frac{1}{2} \sum\limits_{i=1}^n \norm{\v_{\r i}}^2.
\end{align}
We can define the position and velocity vector with respect to the center of mass, i.e. $\mathbf{c} \triangleq [\mathbf{c}_1, ..., \mathbf{c}_n ] \in \R^{3\times n}$, where $\mathbf{c}_i = \r_i - \r_{\textbf{com}}$; $\mathbf{v_c} \triangleq [\v_{\mathbf{c}1}, \v_{\mathbf{c}2}, ..., \v_{\mathbf{c}n} ] \in \R^{3\times n}$, where  $\v_{\mathbf{c} i} = \v_{\r i} - \v_{\textbf{com}}$. Then 
\begin{align}
E_{\text{com}} = \frac{n}{2} \norm{\v_{\textbf{com}}}^2, \quad E_{\text{rel}} \triangleq \frac{1}{2} \sum\limits_{i=1}^n \norm{\v_{\mathbf{c}i}}^2.
\end{align}
We thus have the splitting, $E = E_{\text{com}} + E_{\text{rel}}$. As shown in \cite{mischiati2017geometric}, instantaneous relative energy allocations can be expressed on the probability simplex\footnote{$\Delta^{k-1}$ denotes the $(k-1)$ dimensional probability simplex
\begin{align*}
\Delta^{k-1} = \left\{x = (x_1,x_2,\hdots,x_k) \in \R^k: 0 \leq x_i \leq 1, \sum\limits_{i=1}^{k}x_i=1 \right\}
\end{align*}
}, $\Delta^4$ by exploiting two distinct fiber bundle structures of the flock's total configuration space to split the total kinetic energy using (i) ensemble fibration or (ii) shape fibration. 

\begin{itemize}
\item[(i)] \textit{Ensemble Fibration}: 
We note that the ensemble inertia tensor $K$ \eqref{eq:K} is (for a generic flock configuration) a symmetric positive definite matrix. Hence its eigendecomposition can be written as, $K = Q \Lambda Q^{\mathsf{T}}$, with $\Lambda = \text{diag}(\lambda_1, \lambda_2, \lambda_3)$, where $\lambda_1 \geq \lambda_2 \geq \lambda_3 > 0$. Define, $F := \mathbf{c} \v_{\mathbf{c}}^{\mathsf{T}} + \v_{\mathbf{c}} \mathbf{c}^{\mathsf{T}}$ and $\tilde{F} = [\tilde{F}_{ij}] = Q^{\mathsf{T}} F Q$. Then the following energy modes can be calculated
\begin{align}
\begin{split}
E_{\text{ens.rot}} &\triangleq \frac{1}{2} \left( \frac{\tilde{F}_{12}^2}{\lambda_1 + \lambda_2} + \frac{\tilde{F}_{13}^2}{\lambda_1 + \lambda_3} + \frac{\tilde{F}_{23}^2}{\lambda_2 + \lambda_3} \right) \\ 
E_{\text{ens.def}} &\triangleq \frac{1}{8} \left( \frac{\tilde{F}_{11}^2}{\lambda_1} + \frac{\tilde{F}_{22}^2}{\lambda_2} +\frac{\tilde{F}_{33}^2}{\lambda_3} \right).
\end{split}
\end{align}
Furthermore,
\begin{align}
E_{\text{vol}} \triangleq \frac{1}{2} \frac{\text{tr}^2 \left( \mathbf{c} \v_{\mathbf{c}}^\mathsf{T}\right)}{\text{tr}(K)},
\end{align}
so that, $E_{\text{ens.res}} = E_{\text{ens.def}} - E_{\text{vol}}$. We may also calculate $E_{\text{dem}} = E_{\text{rel}} - E_{\text{ens.rot}} - E_{\text{ens.def}}$. For interpretations of these
energy modes, see \cite{mischiati2017geometric}. Hence, in this fibration we have the following splitting of the kinetic energy
\begin{align}
\left( \frac{E_{\text{com}}}{E}, \frac{E_{\text{dem}}}{E}, \frac{E_{\text{ens.rot}}}{E}, \frac{E_{\text{vol}}}{E}, \frac{E_{\text{ens.res}}}{E} \right) \in \Delta^4.
\label{eq:energy_modes_ensemble}
\end{align}

\item[(ii)] \textit{Shape Fibration}: 
Define the flock angular momentum
\begin{align}
\begin{split}
\mathbf{J} &= \sum\limits_{i=1}^n \left(\mathbf{c}_i \times \v_{\mathbf{c}i} \right), \\
I_\mathbf{c} &= \sum\limits_{i=1}^n \left( \norm{\mathbf{c}_i}^2 \mathbb{1} - \mathbf{c}_i\mathbf{c}_i^\mathsf{T}\right).
\end{split}
\end{align}
Then the rotational energy $E_{\text{rot}}$ can then be calculated as
\begin{align}
E_{\text{rot}} \triangleq \frac{1}{2} \mathbf{J}^\mathsf{T} I_{\mathbf{c}}^{-1} \mathbf{J},
\end{align}
The shape residual energy is given by $E_{\text{shp.res}} = E_{\text{rel}} - E_{\text{rot}} - E_{\text{ens.def}}$, which provides the splitting in this fibration as below
\begin{align}
\left(\frac{E_{\text{com}}}{E}, \frac{E_{\text{rot}}}{E}, \frac{E_{\text{shp.res}}}{E}, \frac{E_{\text{vol}}}{E}, \frac{E_{\text{ens.res}}}{E} \right) \in \Delta^4.
\label{eq:energy_modes_shape}
\end{align}
\end{itemize}

\begin{figure*}[!t]
\centering
	\includegraphics[width=\linewidth, trim = 0 4cm 0 0, clip = true]{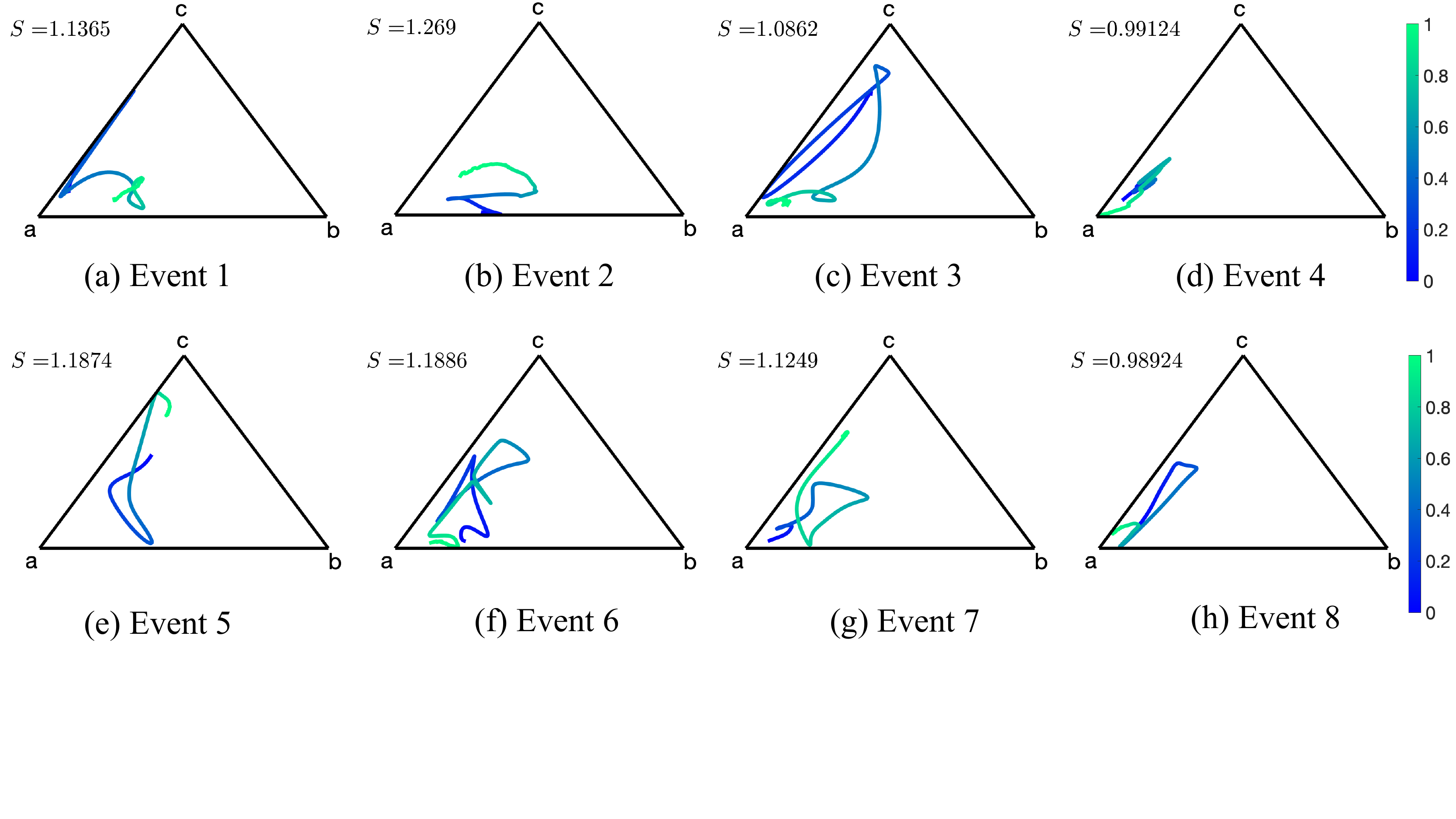}
	\caption{Signatures on the 2-D simplex using ensemble fibering for all eight events. Each signature is colored by normalized time, with initial time in blue and final time in green. Vertices a, b and c of the simplex correspond to the kinetic energy fractions of $E_{\text{dem}}/E_{\text{rel}}$, $E_{\text{ens.rot}}/E_{\text{rel}}$ and $E_{\text{ens.def}}/E_{\text{rel}}$, and $S$ is the entropic cost calculated for each signature in base two. Except for Event 4 which shows little variation in its allocation of $E_{\text{rel}}$ to the democratic strategy, other events display a more complex evolution of the energy distributions to predominantly two of the components.}
	\label{fig:2d_ens}
\end{figure*}

While we can split the kinetic energy in 5 different modes \eqref{eq:energy_modes_ensemble},\eqref{eq:energy_modes_shape}, many flocking events show a predominant allocation of nearly constant energy of rigid translation ($E_{\text{com}}$) (see Appendix~\ref{appdx:suppl}). We exclude this component from the total $E$ in our analysis, and consider the allocation of the remaining energy $E_{\text{rel}}$ to obtain a time dependent signature of each event on a lower dimensional simplex. In particular, we capture the signature generated by the following decomposition of $E_{\text{rel}}$ using ensemble fibration on the 1-simplex by two different methods
\begin{flalign}
(\text{ENS-I})  \hspace{4.5cm}	\left( \frac{E_{\text{dem}}}{E_{\text{rel}}}, \frac{E_{\text{ens}}}{E_{\text{rel}}} \right) \in \Delta^1, && 
\label{eq:game1}
\end{flalign}
\begin{flalign}
(\text{ENS-II}) \hspace{4.4cm}	\left( \frac{E_{\text{ens.rot}}}{E_{\text{rel}}}, \frac{E_{\text{rel}} - E_{\text{ens.rot}}}{E_{\text{rel}}} \right) \in \Delta^1, &&
\label{eq:game2}
\end{flalign}
where $E_{\text{rel}} = E - E_{\text{com}}$, and $E_{\text{ens}} = E_{\text{rel}} - E_{\text{dem}} = E_{\text{ens.rot}} + E_{\text{vol}} + E_{\text{ens.res}}$. 
Similarly, a one dimensional simplex description using shape fibration may be given by two ways
\begin{flalign}
(\text{SHP-I})  \hspace{4.5cm}	\left( \frac{E_{\text{shp.res}}}{E_{\text{rel}}}, \frac{E_{\text{rel}}-  E_{\text{shp.res}}}{E_{\text{rel}}} \right) \in \Delta^1, && 
\label{eq:game3}
\end{flalign}
\begin{flalign}
(\text{SHP-II}) \hspace{4.4cm}	\left( \frac{E_{\text{rot}}}{E_{\text{rel}}}, \frac{E_{\text{shp}}}{E_{\text{rel}}} \right) \in \Delta^1, &&
\label{eq:game4}
\end{flalign}
where $E_{\text{shp}} = E_{\text{rel}} - E_{\text{rot}} = E_{\text{shp.res}} + E_{\text{vol}} + E_{\text{ens.res}}$.
\begin{figure*}[!t]
\centering
	\includegraphics[width=\linewidth, trim = 0 4cm 0 0, clip = true]{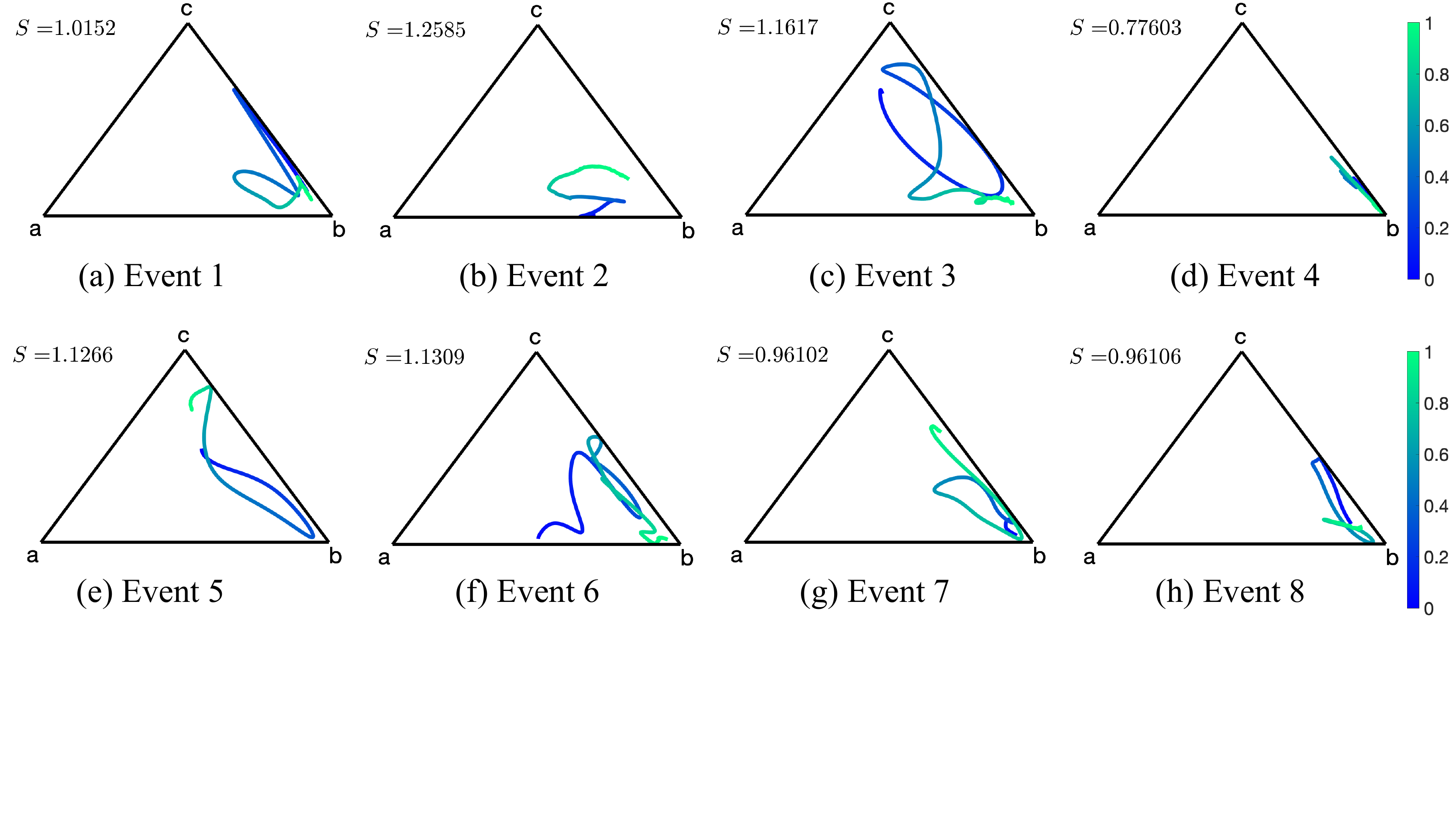}
	\caption{Signatures on the 2-D simplex using shape fibering for all eight events. Each signature is colored by normalized time, with initialized time in blue and final time in green. Vertices a, b and c of the simplex correspond to the kinetic energy fractions of $E_{\text{rot}}/E_{\text{rel}}$, $E_{\text{shp.res}}/E_{\text{rel}}$ and $E_{\text{ens.def}}/E_{\text{rel}}$, and $S$ is the entropic cost calculated for each signature in base two. Similar to the signatures corresponding to the ensemble fibering, Event 4 has the least variation in the allocation of $E_{\text{rel}}$ almost entirely to $E_{\text{shp.res}}$ over its total duration, while other events undergo more complex evolution of the energy distributions.}
	\label{fig:2d_shp}
\end{figure*}
On the 1-dimensional simplex we have four different geometric ways of looking at flocking event signatures. On the 2-dimensional simplex $\Delta^{2}$ one has two different signatures for each flocking event 
\begin{flalign}
(\text{2DENS}) \hspace{4cm} \left(\dfrac{E_{\text{dem}}}{E_{\text{rel}}}, \dfrac{E_{\text{ens.rot}}}{E_{\text{rel}}}, \dfrac{E_{\text{ens.def}}}{E_{\text{rel}}}\right) \in {\Delta^2}, && 
\end{flalign}
\begin{flalign}
(\text{2DSHP}) \hspace{4cm} \left(\dfrac{E_{\text{rot}}}{E_{\text{rel}}}, \dfrac{E_{\text{shp.res}}}{E_{\text{rel}}}, \dfrac{E_{\text{ens.def}}}{E_{\text{rel}}}\right) \in {\Delta^2}, &&
\end{flalign}
corresponding respectively to the ensemble and shape fibrations. Letting~$E_i, \ i=1,2,3$ stand in for fractional energy modes, one can associate an average entropy $S =\langle -\sum_{i=1}^{3}E_i \log_2 E_i \rangle$ to each event (the average denoted by angle bracket $\langle \ \cdot \ \rangle$ is taken over the duration of the event). The results are shown in Fig.~\ref{fig:2d_ens} and Fig.~\ref{fig:2d_shp}. We will refer to $S$ as the \textit{entropic cost}. For both SHP and ENS fiberings Event 2 has the highest entropic cost, although event-to-event variability in this measure is not very substantial. 

From the formulas relating the different energy modes it follows that one can view the $\Delta^1$ signature as a compression of the $\Delta^2$ signature. One should note that entropic cost is not sensitive to temporal variations in the signature. By fitting suitable dynamic models to signatures, one can develop a measure that is sensitive to temporal variability. We do precisely this in Section~\ref{sec:generative}, working out the details in the setting of signatures on $\Delta^1$. As we shall see, this also reveals fibering dependencies of signature properties. Our approach is based on generative evolutionary game dynamics modeling the competition between the flock-scale strategies.

The moment-to-moment decisions made by individuals in a flock, taking account of the decisions by their conspecific neighbors and the predators, contribute to flock-scale strategies captured in the present section  by time dependent signatures on the probability simplex. Fractional energy modes are conceptualized as probabilities defining mixed strategies in a game. This sets the stage for an evolutionary game-theoretic treatment of flocking events and quantitative comparisons of events.


\section{Generative Models on the Simplex and the Data-smoothing Problem} 
\label{sec:generative}
\begin{figure*}[t]
\centering
	
		\includegraphics[scale=0.35]{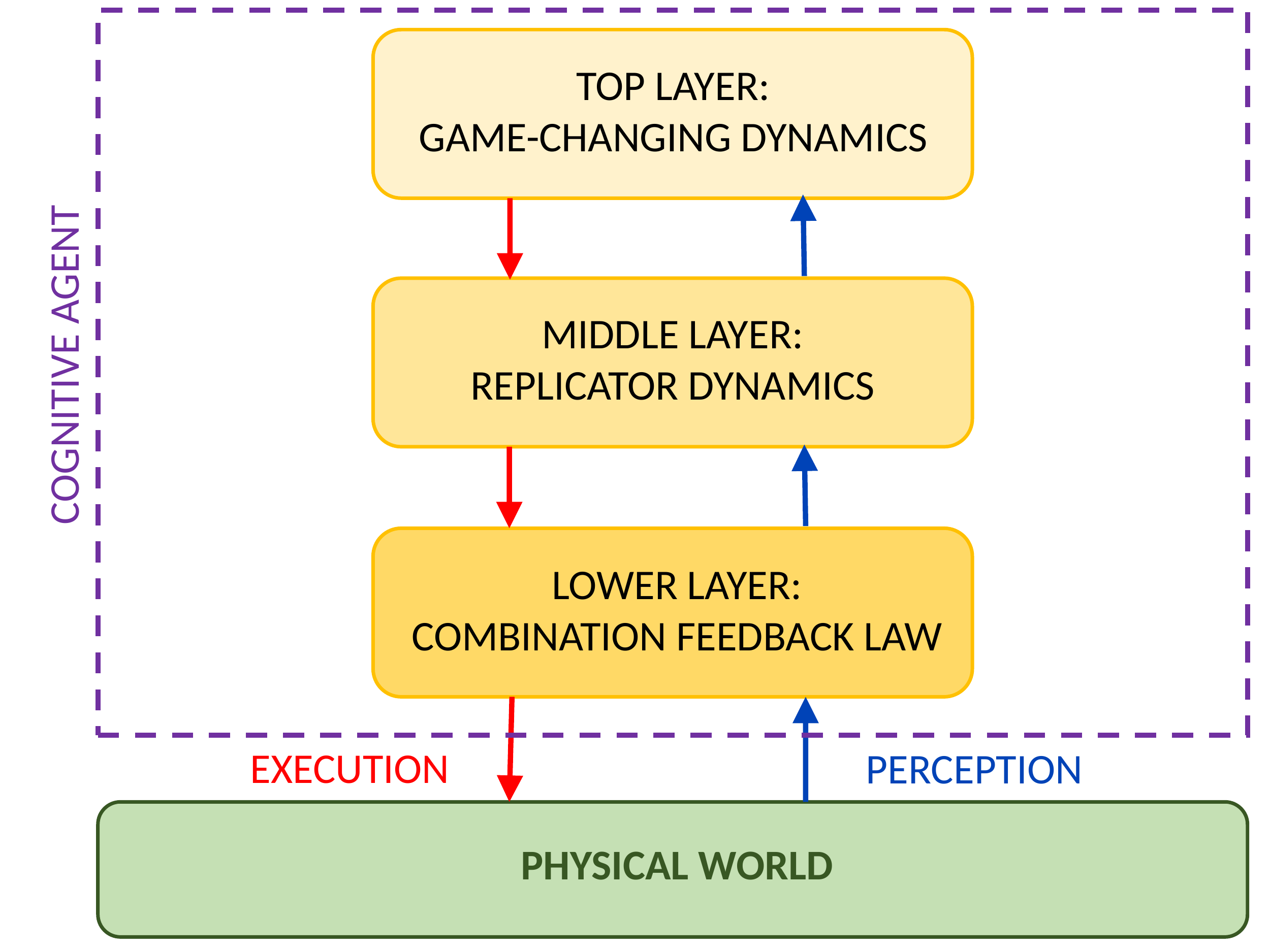}
	
	\caption{The cognitive hierarchy}
	\label{fig:cog}
\end{figure*}
In this work, instead of modeling the microscopic behavior of individual members of the flock that result in complex collective motion, motivated by \cite{mischiati2017geometric}, we adopt the viewpoint that the flock operates \textit{as if} it is a single cognitive entity, capable of exhibiting different `modes' of behavior. These modes or strategies are identified with the allocation of kinetic energy components. The temporal variations of a resulting signature on the probability simplex as explained in Section~\ref{sec:energymodes} can be captured by an underlying controlled evolutionary game that models competition between strategies. The average Hamiltonian associated to optimal controls in this setting is interpreted as a measure of cognitive cost.   


Our approach in this section is informed by the concept of a three-layered cognitive hierarchy operating at multiple time-scales, shown in Fig. \ref{fig:cog}, to model decision-making by the collective \cite{raju2019cognitive}. The bottom layer captures the interaction of the flock with its environment at a fast time-scale. This interaction is in accordance with a mixed-strategy choice determining the allocations of kinetic energy, dictated by replicator dynamics in the middle layer operating at a slower time-scale. The fitness map defining this replicator dynamics is in turn modulated by means of controls in the top layer, at the slowest time-scale. While the time-scale differences of this model are meant to distinguish reaction to fast external stimuli from long term learning, the top layer can flexibly intervene in order to accommodate for a changing environment or enable response to an adversary. With these in mind, we propose a class of generative models on the simplex. 

Controlled evolutionary games offer a natural model for capturing the underlying dynamics that generates signatures representing time-dependent mixed strategies on a simplex. The replicator dynamics that captures the evolutions of fractions $x_i$ of $k$ types is given by
\begin{align}
\dot{x} = \Lambda(x)(f(x)- \bar{f}(x)\textbf{e}) \label{repDyn},
\end{align}
where $\Lambda(x) = ~\text{diag}(x_1, x_2, \cdots, x_k) \in \R^{k\times k}$, $f(x) = [f^1(x) \ \hdots \ f^k(x)]^\mathsf{T} \in \R^k $ is a fitness map such that each component is an element of $C^\infty(\Delta^{k-1})$, the space of smooth functions on the simplex, $\bar{f}(x) = \sum\limits_{i=1}^k x_i f^i(x)$ is the average fitness, and $\textbf{e} = [1 \ \hdots \ 1]^\mathsf{T} \in \R^k$. Replicator dynamics have been shown to be universal in recent work \cite{raju2019lie}. That is, every simplex-preserving dynamics can be transformed into replicator dynamics with appropriate fitness. Therefore, they are natural candidates for generative models on the simplex, as explained below. 

Consider the system on $\Delta^{k-1}$ \cite{raju2019lie}
\begin{align}
\dot{x} &= u_1\hat{f}_1 + u_2 \hat{f}_2, \label{eq:genDyn}
\end{align}
where $f_i, i=1,2$ are fitness maps, and $\hat{f}_i, i=1,2$ are associated replicator vector fields, and $u_i, i=1,2$ are controls. Suppose the fitness maps are given by $f_1 = [a_1 \hdots a_k]^\mathsf{T} \in \R^k$, a frequency independent fitness, and $f_2 = Bx$, a linear fitness map where all $a_i$'s are assumed to be distinct and positive, and $B$ is assumed to be non-singular. Then, according to \cite{raju2019cognitive, raju2019lie}, (\ref{eq:genDyn}) is controllable. This implies that there exist controls $u_i, i=1,2$ such that any two points in the interior of the probability simplex can be connected by a solution curve of (\ref{eq:genDyn}). The dynamics (\ref{eq:genDyn}) with the specific choice of fitness maps can be used to explain/approximate signatures on the simplex for two reasons: first, due to controllability, and second, such a system allows us to model the competition between the strategies for a cognitive agent offering interpretability. When $u_2 = 0$ and $u_1 = 1$ identically, the behavior of (\ref{eq:genDyn}) is to converge to a pure strategy. For this reason, $\hat{f}_1$ can be identified as dynamics due to a bias contributed by the ordering of the frequency independent fitness components learned via games against nature as in \cite{wei2009pursuit}. On the other hand, when $u_1= 0$ and $u_2 = 1$ identically, the evolution of the strategies is influenced by the game matrix $B$ which reflects a comparative assessment of the pure strategies when pitted against each other. Therefore, (\ref{eq:genDyn}) is interpreted to be a system capable of producing any desired mixed strategy decision, by managing the influence of pre-existing cognitive biases and learned information or experience, with the controls as driving forces.

\subsection{Specialization to $k=2$}
\label{sec:generative_special2}
Since we are interested in describing the evolution of two flock strategies as in \cref{eq:game1,eq:game2} for ensemble fibration or \cref{eq:game3,eq:game4} for shape fibration, we capture the signature of a flocking event via a generative model on the 1-simplex. We consider an evolutionary game model, namely replicator dynamics equipped with a multiplicative control, in order to describe their evolution in the interior $(0,1)$ of the one-dimensional simplex. The choice of replicator dynamics is influenced by its universality in describing simplex-preserving dynamics, and by virtue of determining extremals for a variational problem \cite{svirezhev1972optimum,raju2018variational}. With the inclusion of a control variable, we consider a different variational problem that aims to perform data smoothing using regularization as in \cite{dey2014control}. To see this, let $\underline{x} = [x_1 \ x_2]^{\mathsf{T}} \in \Delta^1$ where $x_i$, $i=1,2$ denote the prevalence of strategies $i$ (to be specified) on the simplex with the natural constraint $x_1+x_2=1$. $x_i = 1, i=1,2$ correspond to allocation of $E_{\text{rel}}$ entirely to one of the two pure strategies.   Suppose that the frequencies associated with the strategies are updated according to the rule
\begin{align}
x_i(t+1) &= x_i(t)\dfrac{f^i(\underline{x})}{\bar{f}(\underline{x})}, \label{discdyn}
\end{align}
where the fitness map $f(\underline{x}) = A\underline{x}$ and $\bar{f}(\underline{x}) = x_1f^1(\underline{x})+x_2 f^2(\underline{x})$. Here, $A = [a_{ij}]$ defines a payoff matrix with $a_{ij}$ denoting the payoff of the $i^{\text{th}}$ strategy against $j^{\text{th}}$ strategy. (In the case that the payoffs do not depend on the strategy $j$ of against which it is matched up, the columns of $A$ are identical.) In the ode limit of (\ref{discdyn}), after an inhomogeneous time-scale change, we get the equations
\begin{align}
\dot{x}_i (t) = x_i(t) (f^i(\underline{x}) - \bar{f}(\underline{x})), ~~i=1,2. \label{eq:xi_dot}
\end{align}
It can be readily verified that (\ref{eq:xi_dot}) is simplex-preserving, leaving the pure strategies invariant. Since addition of the same term to each component of the fitness keeps the dynamics (\ref{eq:xi_dot}) unchanged, by subtracting $a_{21}$ and $a_{12}$ from the first and second column elements of $A$ respectively, we get the equivalent payoff matrix
\begin{align}
\tilde{A} &= \left[\begin{array}{cc}
a_{11}-a_{21} & 0 \\
0 & a_{22}-a_{12}
\end{array}\right]. 
\end{align}
We restrict the parameters of the matrix such that $a_{11}-a_{21}  = -( a_{22}-a_{12})=\beta$ so that the fitness can be rewritten as
\begin{align}
f(\underline{x}) &= \beta\left[\begin{array}{cc}
1 & 0 \\
0 & -1
\end{array}\right]\underline{x}.
\end{align}
Due to the simplex constraint, (\ref{eq:xi_dot}) is completely described using $x=x_1$,
\begin{align}
\dot{x} (t) =  \beta x(t) (1-x(t)),
\label{eq:x_dot}
\end{align}
with $x=0,1$ corresponding to the pure strategies 2 and 1 respectively. This allows us to adopt a time-scale change by the factor $\beta$ and introduce a time-varying control to modulate the fitness as in (\ref{eq:genDyn}) to arrive at our generative model 
\begin{align}
\dot{x} (t) = u(t) x(t) (1-x(t)).
\label{eq:x_dot_generative}
\end{align}
This dynamics results in asymptotic convergence to the pure strategy $x=1$ in the absence of control, that is, when $u(t) \equiv 1$. However, the time-varying control variable $u$ serves to model changing preferences for the flock strategies by appropriate changes in its sign and magnitude. Such a temporal modulation of the fitness ensures feasibility of capturing arbitrary signatures in the interior of the simplex. \\

\subsection{Optimal Control Problem}
\label{sec:generative_optimal}
Given a set of data points $\{x_0^d, x_1^d, ..., x_N^d \}$ with each $x_k^d \in (0,1), k = 0, 1, ..., N$, at time instants $\{t_0, t_1, ..., t_N\}$, we formulate the optimal control problem
\begin{align}
\begin{split}
& \min\limits_{x(t_0), ~ u(\cdot)} J(x(t_0), u) = \sum\limits_{i=0}^N F_i (x(t_i)) + \frac{\lambda}{2}\int_{t_0}^{t_N} u^2(t) \, dt ,  \\ 
& \text{subject to:~~} \dot{x} = u x(1-x), 
\end{split}
\label{eq:optimal_control_problem}
\end{align}
where the fit errors $F_i$'s are given by the Kullback-Leibler divergence measure of mismatch between the data and the state
\begin{align}
F_i(x) = x_i^d \log \left( \frac{x_i^d}{x} \right) + (1 - x_i^d) \log \left( \frac{1-x_i^d}{1-x} \right), \quad i = 0, 1, ..., N.
\label{eq:KL_divergence}
\end{align}
We can directly appeal to Pontryagin's Maximum Principle (PMP) and theorem (\ref{thm:PMP_dataSmoothing}) to write necessary conditions for optimality.
We can write the pre-Hamiltonian as
\begin{align}
H(x, p, u) = u p x (1-x) - \frac{\lambda}{2} u^2.
\end{align}
The Hamiltonian maximization condition \eqref{eq:hamiltonian_max} yields an optimal control in each time interval $t \in (t_i, t_{t+1}), ~ i = 0, 1, ..., N-1$,
\begin{align}
u = \frac{1}{\lambda} p x(1-x),
\end{align}
with the Hamiltonian given by
\begin{align}
\H (x, p) = \frac{1}{2\lambda} p^2 x^2(1-x)^2.
\label{eq:hamiltonian}
\end{align}
Hamilton's equations \eqref{eq:xp_dot_thm} read
\begin{align}
\begin{split}
\dot{x} &=  \frac{1}{\lambda} p x^2 (1-x)^2 \\
\dot{p} &= -\frac{1}{\lambda} p^2 x (1-x) (1-2x).
\end{split}
\label{eq:xp_dot}
\end{align}
The jump conditions for $p$ \eqref{eq:jump_p_thm} can be written as
\begin{align}
\begin{split}
& p(t_0^-) = 0, \\
& p(t_i^+) - p(t_i^-) = \frac{x(t_i) - x_i^d}{x(t_i) (1-x(t_i))}, \quad i = 0, 1, ..., N, \\
& p(t_N^+) = 0.
\end{split}
\label{eq:jump_p}
\end{align}
\begin{remark}
Note that the optimal control is piecewise constant since $\tfrac{du}{dt} = 0$ for each of these time intervals $t \in (t_i, t_{i+1}), ~ i = 0, 1, ..., N-1$. 
\end{remark}
Therefore, denoting $x_k = x(t_k), ~ k = 0, 1, ..., N$, any optimal control can be described by a tuple $(u_0, u_1, ..., u_N)$ with the conditions
\begin{align}
\begin{split}
& u_0 = \frac{1}{\lambda} (x_0 - x_0^d), \\
& u_k - u_{k-1} = \frac{1}{\lambda} (x_k - x_k^d), \quad k = 1, 2, ..., N, \\
& u_N = 0.
\end{split}
\label{eq:jump_u}
\end{align}
Piecewise constancy of the control input allows us to write the solution to the state equation \eqref{eq:x_dot_generative} explicitly. Suppose the sampling time of the signature is uniform, i.e. $\Delta t := t_{k+1} - t_k, \forall k \in \{0, ..., N-1\}$, integrating the state equation \eqref{eq:x_dot_generative} in $(t_k, t_{k+1})$, we can write
\begin{align}
x_{k+1} = \frac{x_k e^{u_k \Delta t}}{1 + x_k \left( e^{u_k \Delta t } -1 \right)}, \quad k = 0, 1, ..., N-1.
\end{align}
By iteration, we can in turn write every $x_k$ as a function of $x_0$ and $u_0, u_1, ..., u_{k-1}$,
\begin{align}
x_k = x_k(x_0) = \frac{x_0 e^{\left(u_0 + u_1 + \cdots + u_{k-1}\right) \Delta t}}{1 + x_0 \left( e^{\left(u_0 + u_1 + \cdots + u_{k-1}\right) \Delta t} -1 \right)}, \quad k = 1, 2, ..., N.
\label{eq:x_k}
\end{align}
The endpoint condition ($u_N = 0$) can then be written as
\begin{align}
g(x_0) = \sum\limits_{k=0}^N (x_k - x_k^d) = 0
\label{eq:endpoint_condition}
\end{align}
where we use \eqref{eq:x_k}.
Solving the optimal control problem \eqref{eq:optimal_control_problem} thus boils down to solving \eqref{eq:endpoint_condition} for $x_0 \in (0,1)$, subject to \eqref{eq:x_k}. 

\begin{figure*}[!h]
\centering
	\includegraphics[width=\textwidth, trim = 0 0 0 0, clip = false]{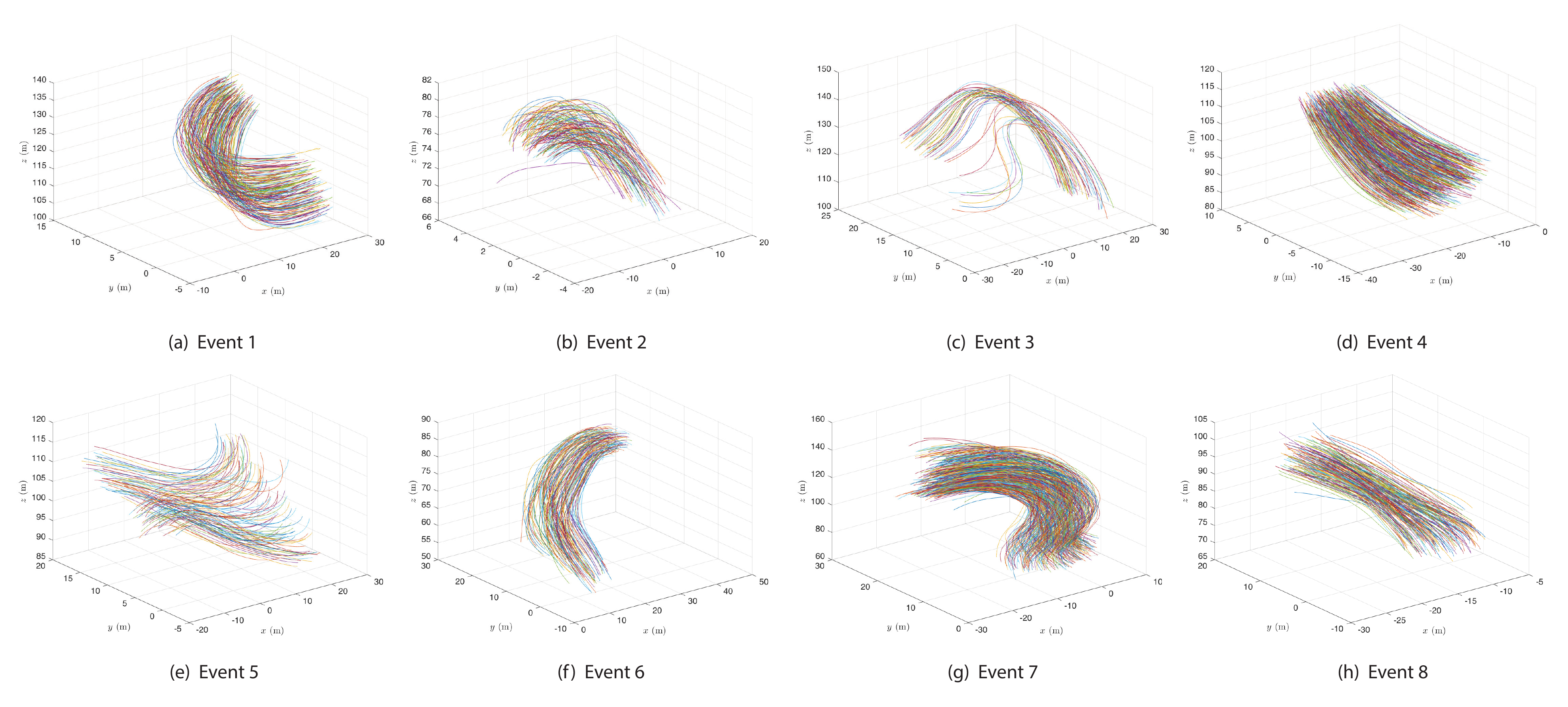}
	\caption{Starling flight trajectories for all eight events. The time-sampled raw data was processed to generate smooth trajectories as shown here. The details for each event are given in Table \ref{tbl:flock}.}
	\label{fig:flight_trajectories}
\end{figure*}

\begin{figure*}[!h]
\centering
	\includegraphics[width=\textwidth, trim = 0 0 0 0, clip = false]{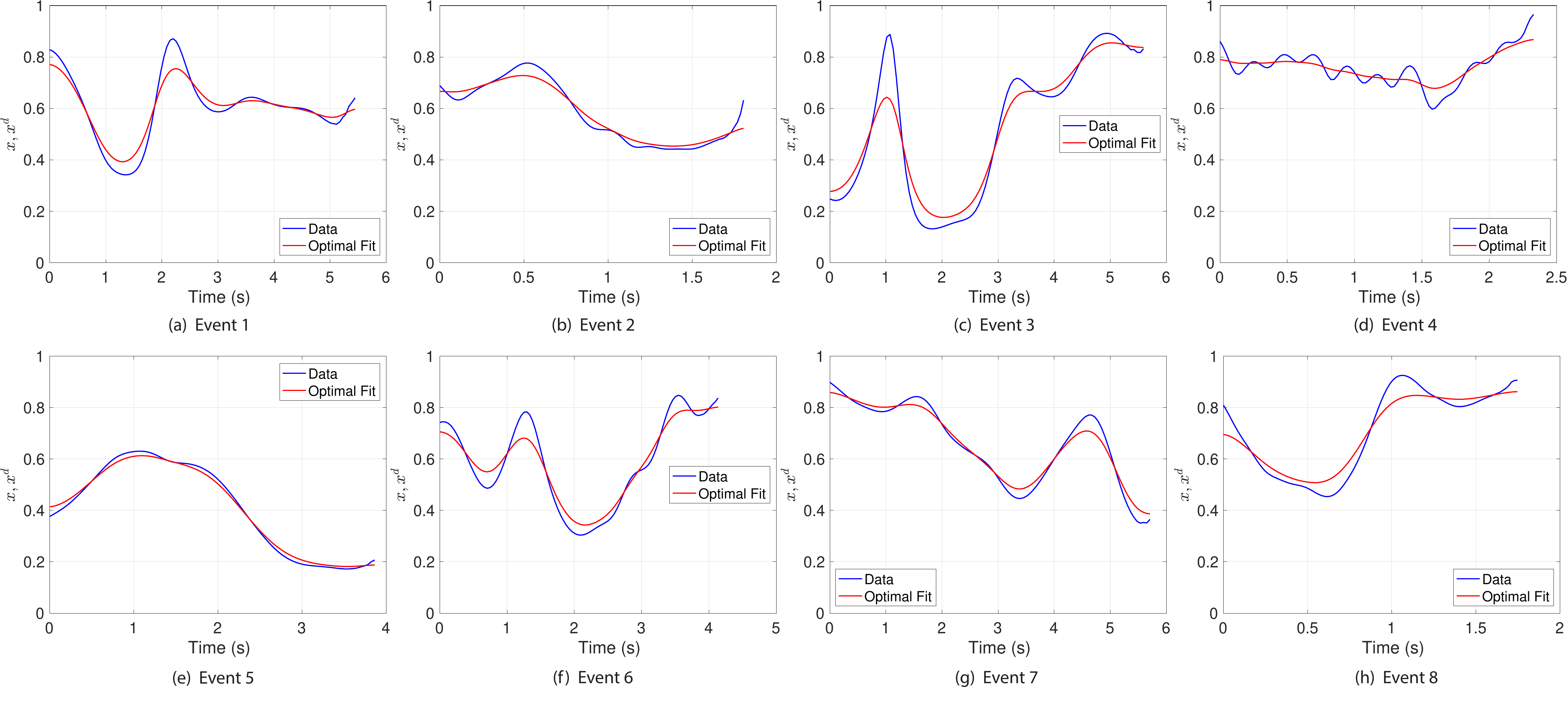}
	\caption{Calculated energy signature and its optimal fit on the 1-D simplex, for the case  ENS-I, i.e. here $x = \tfrac{E_{\text{dem}}}{E_{\text{rel}}}$.}
	\label{fig:fit_ens1}
\end{figure*}

\begin{figure*}[!ht]
\centering
	\includegraphics[width=\textwidth, trim = 0 0 0 0, clip = false]{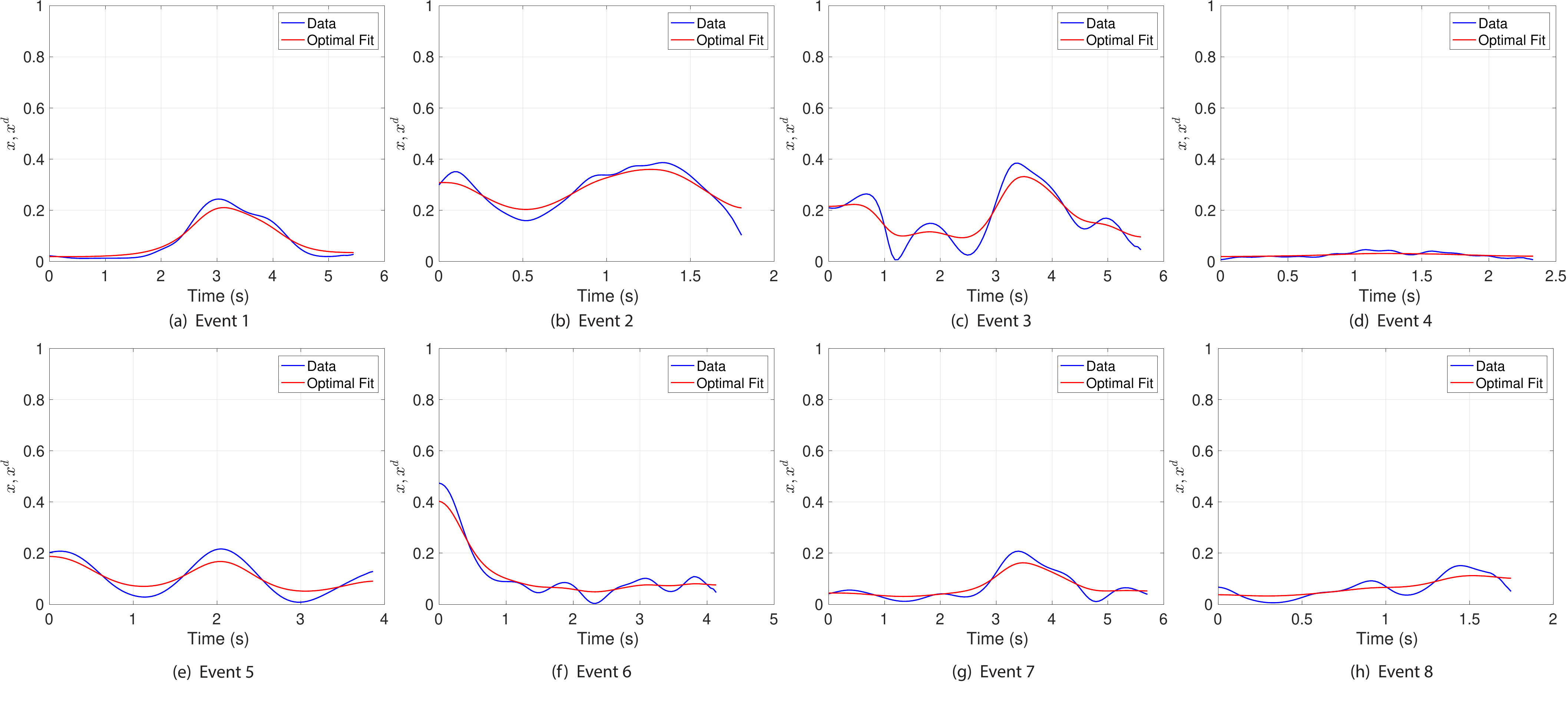}
	\caption{Calculated energy signature and its optimal fit on the 1-D simplex, for the case  SHP-II, i.e. here $x = \tfrac{E_{\text{rot}}}{E_{\text{rel}}}$.}
	\label{fig:fit_shp2}
\end{figure*}

\begin{remark}
The value of the regularization parameter $\lambda$ is usually chosen through cross validation technique. We do not employ this. The value of $\lambda$ is chosen such that the root finding algorithm for solving \eqref{eq:endpoint_condition} converges for all events. For $\lambda = 0.2$, (assuming the logarithms in equation \eqref{eq:KL_divergence} are natural logarithms) the roots were found with reasonably good accuracy 
for all events with the value of the function $g(\cdot)$ dropping to the order of $10^{-5}$. For lower $\lambda$ however, the problem becomes stiffer and left hand side of \eqref{eq:endpoint_condition} demonstrates `effective discontinuity' in $x_0$. This poses serious problem in solving \eqref{eq:endpoint_condition}. As a future step, cross validation could be employed to arrive at a good value of $\lambda$ in the range where \eqref{eq:endpoint_condition} can be solved.
\end{remark}

\begin{remark}
A concern with not performing cross validation to obtain the parameter $\lambda$ is overfitting to noise. We note here that the original flight data was subjected to data-smoothing to obtain smooth trajectories \cite{dey2015reconstruction}. This data-smoothing process used ordinary cross validation for trajectory of each bird to determine the appropriate weight to the regularization term. This procedure generated smooth trajectories with suppressed level of noise compared to the original data. We then construct the sampled signature $\{x_0^d, \cdots, x_N^d \}$ from these smooth trajectories. Therefore, not performing cross validation for the data smoothing on the simplex may not be as restricting as
one might expect.
\end{remark}


\section{Signature Fitting Results and Cognitive Cost} \label{sec:results}

For all eight events, we perform the data smoothing technique as described in Section \ref{sec:smoothing_linear} to obtain smooth flight trajectories. These are shown in Fig. \ref{fig:flight_trajectories}. Given the smooth trajectories for all the birds in a flock, we compute signatures in $\Delta^1$ and solve the optimal control problem \eqref{eq:optimal_control_problem} and report the results here.  The value of the regularization weight $\lambda$ is taken to be 0.2 and 100 signature data samples at regular time intervals are taken for all events. Given this data vector, we solve equation \eqref{eq:endpoint_condition}, constrained by \eqref{eq:x_k}, for $x_0 \in (0, 1)$. Optimal control solutions for the \textit{games} ENS-I \eqref{eq:game1} and SHP-II \eqref{eq:game4} for individual events are shown in Fig. \ref{fig:fit_ens1} and \ref{fig:fit_shp2}, respectively. In Table \ref{tbl:results}, we report time-averaged Hamiltonian integrals or simply the average Hamiltonian and time-averaged total costs for all the different games that we consider in \cref{eq:game1,eq:game2,eq:game3,eq:game4}. 

\begin{figure*}[t]
\centering
	\includegraphics[width=\textwidth, trim = 0 0 30pt 0, clip = true]{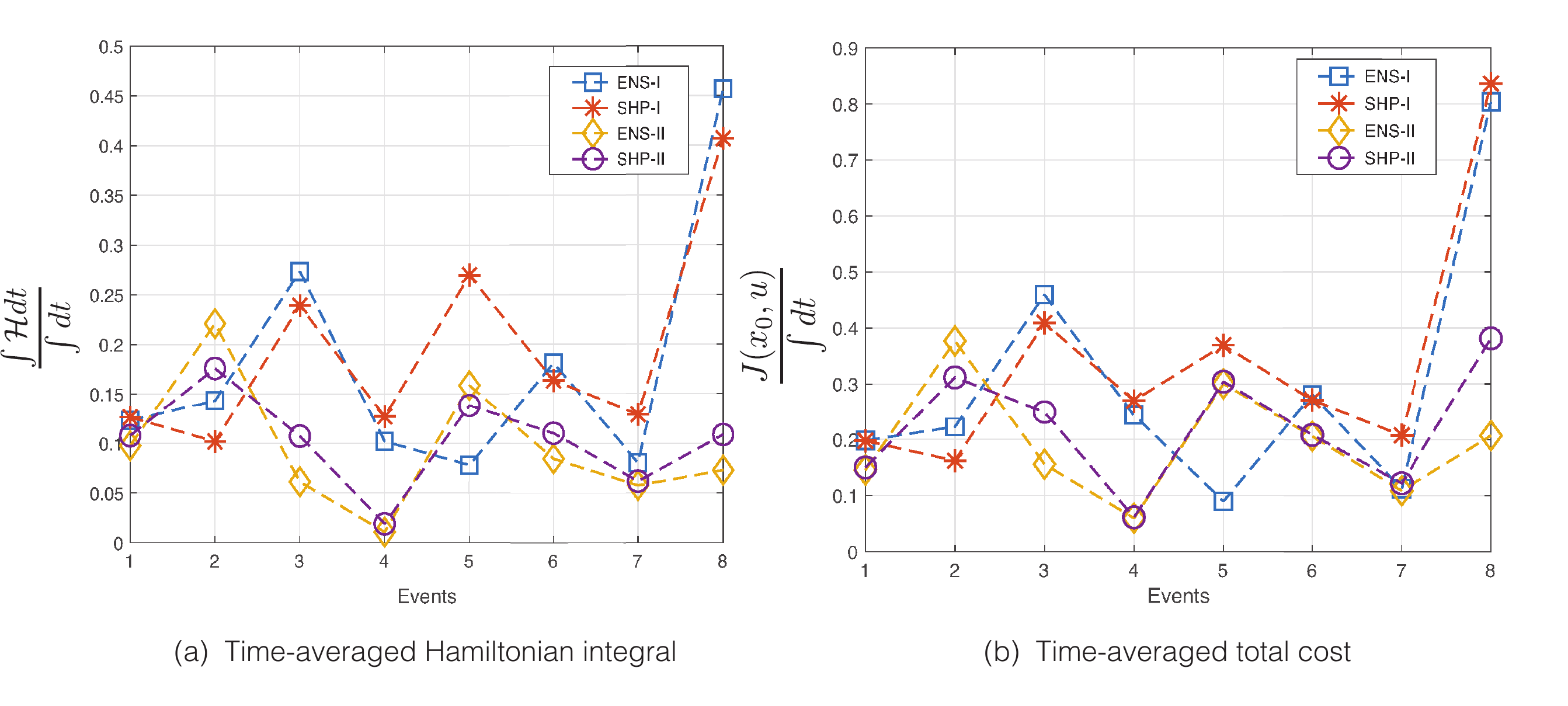}
	\caption{Hamiltonian signatures for all eight flocking events: (a) time-averaged Hamiltonian integral or the cognitive cost, (b) time-averaged optimal cost from the data-fitting problem.}
	\label{fig:hamiltonian_signatures}
\end{figure*}

With the understanding that flocks act in a way that reduces the cognitive burden on individual birds, we interpret the average Hamiltonian as \textit{cognitive cost} of an event. We additionally compute the average optimal cost $J/T$ which includes the effect of the regularization parameter $\lambda$. Both are graphically represented in Fig. \ref{fig:hamiltonian_signatures}. As seen from Table \ref{tbl:results} and Fig.~\ref{fig:hamiltonian_signatures}, the trend of (ENS-I) closely follows that of the game (SHP-I), except for Event 5, while the games ENS-II and SHP-II show similar trend. We note that greater temporal variation in the energy signal results in higher cognitive cost (in both measures). It is \textit{as if} the collective, in rapidly re-allocating energy modes, has to ‘think’ more and thereby incur higher cognitive cost.
These cognitive costs for a particular game can thus indicate overall physical behavior of the flock. For example, in the games (ENS-II) or (SHP-II) where a rotational energy is considered as one of the pure strategies, relatively higher cognitive costs for Events 2 and 5 indicate that the flocks went through more rotations than the other events during the flight periods. 
On the other hand, the low cognitive cost of Event 4 may be explained by noting that the signature on $\Delta^2$ (see Fig.~\ref{fig:2d_ens}) remains close to vertex a i.e. $E_{\text{rel}}$ is nearly all $E_{\text{dem}}$.
Similar conclusions can be drawn for the other set of games (ENS-I) and (SHP-I), where the respective cognitive cost will stipulate nature of variation of the democratic (reshuffling within the flock) energy. The higher the cost, the more aggressively the relative positions of the birds within the flocks are changed, leading to a more complex flight event. This is especially seen in Events 3 and 8.

\begin{table}[!t]
\footnotesize
\centering
  \begin{tabular}{ c | c  c  c  c | c  c  c  c }
     {\cellcolor{black} \color{white}Duration}   & \multicolumn{4}{c|}{\cellcolor{gray!70}  $\frac{\int \H dt}{\int dt}$}  &  \multicolumn{4}{c}{ \cellcolor{gray!70} $\frac{J(x_0, u)}{\int dt}$} \\ 
   {\cellcolor{black} \color{white}(seconds)}  & {\cellcolor{gray!40}}\footnotesize{(ENS-I)} & {\cellcolor{gray!40}}\footnotesize{(SHP-I)} & {\cellcolor{gray!40}}\footnotesize{(ENS-II)} & {\cellcolor{gray!40}}\footnotesize{(SHP-II)} & {\cellcolor{gray!40}}\footnotesize{(ENS-I)} & {\cellcolor{gray!40}}\footnotesize{(SHP-I)} & {\cellcolor{gray!40}}\footnotesize{(ENS-II)} & {\cellcolor{gray!40}}\footnotesize{(SHP-II)} \\
	&&&&&&&& \\ [-0.4em]
    5.4875 & 0.1232 & 0.1263 & 0.0976 & 0.1077 & 0.1981 &  0.1975 & 0.1454 &  0.1499\\ [0.3em] 
    1.8176 & 0.1432 & 0.1018 & 0.2210 & 0.1760 & 0.2227 &  0.1619 & 0.3769 &  0.3118 \\ [0.3em] 
    5.6118 & 0.2735 & 0.2392 & 0.0613 & 0.1073 & 0.4595 &  0.4092 & 0.1557 &  0.2495\\ [0.3em] 
    2.3471 & 0.1021 & 0.1270 & 0.0107 & 0.0190 & 0.2440 &  0.2702 & 0.0594 &  0.0610\\ [0.3em] 
    3.8824 & 0.0779 & 0.2699 & 0.1587 & 0.1383 & 0.0896 &  0.3692 & 0.3001 &  0.3041 \\ [0.3em] 
    4.1588 & 0.1809 & 0.1634 & 0.0846 & 0.1105 & 0.2799 &  0.2706 & 0.2063 &  0.2090\\ [0.3em] 
    5.7353 & 0.0804 & 0.1293 & 0.0576 & 0.0619 & 0.1127 &  0.2079 & 0.1087 &  0.1221\\ [0.3em] 
    1.7588 & 0.4569 & 0.4069 & 0.0731 & 0.1090 & 0.8037 &  0.8361 & 0.2074 &  0.3810\\ [0.3em] \hline  	
  \end{tabular}
  \caption{Costs for all eight flocking events}
  \label{tbl:results}
\end{table}

These observations are reflected in the optimal control signals as well which are presented in Appendix~\ref{appdx:suppl}. For those events with more pronounced strategy changes, in terms of the rate of change of energy allocations, the control effort is higher in magnitude. This suggests what one might expect: more the change in the strategy dictating the energy allocations, higher the cognitive effort. The use of the average Hamiltonian as a cognitive cost captures this understanding. 

\begin{figure*}[th]
    \centering
    \includegraphics[scale = 0.6, trim = 0 10pt 0 0, clip = true]{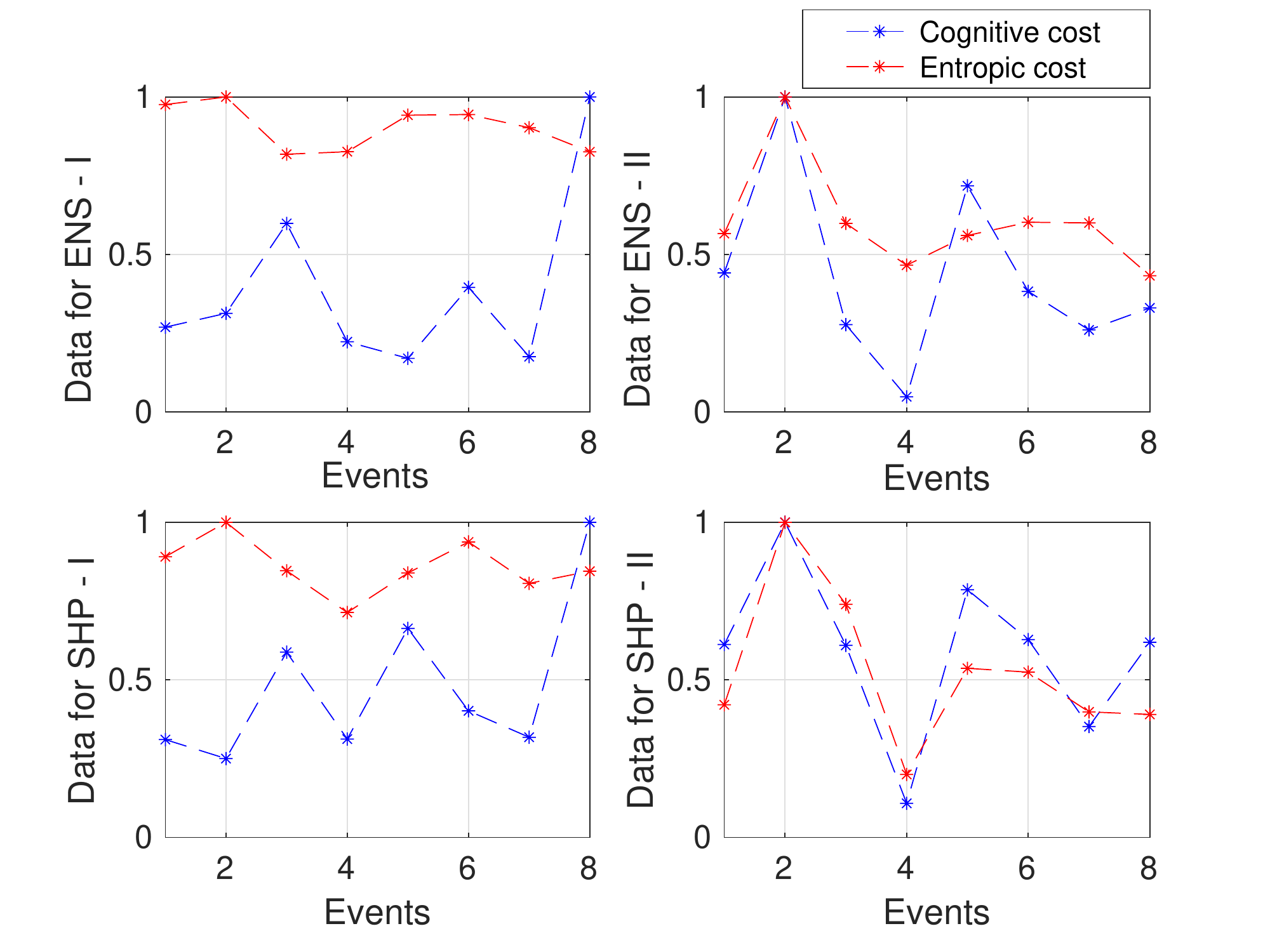}
    \caption{Plot of normalized (by maximum across all eight events) cognitive cost (blue) against entropic cost (red) for all four games.
}
    \label{fig:entropy_all}
    \vspace*{-10pt}
\end{figure*}

As already discussed in Section~\ref{sec:energymodes} in the setting of the 2-dimensional simplex, where we associate an entropic cost $S$, the \textit{empirical} entropic cost for the setting of the 1-dimensional simplex can be computed as 
\begin{align}
 S  &= \frac{1}{N}\left(-\sum_{i=0}^{N}\left(x_i^d \log_2 x_i^d  + \left(1-x_i^d\right) \log_2 \left(1-x_i^d\right) \right) \right) 
\end{align}
The plots of normalized cognitive cost obtained from the smoothed trajectory on the simplex and time-averaged entropy based on the raw signatures on the simplex for all eight events are depicted in Fig.~\ref{fig:entropy_all}. We observe that for ENS-II and SHP-II the entropic cost and the cognitive cost appear to show consistent trends.

\section{Discussion}
\label{sec:discussion}
In this paper, we associate the dynamic behavior of a collective with a cognitive process that realizes
decision-making with benefits to agents in the collective. The process is envisioned as distributed across
the collective and not embodied in any one agent. In the setting of starling flocks this is achieved by
flock-scale behaviors (emerging from individuals interacting with conspecific individuals and predators),
recognizable as morphological changes over time, and quantified as dynamic allocations of kinetic
energy modes. The different energy modes are thought of as pure strategies of a controlled evolutionary
game with the fitness modulated by time-varying decision or control variables. From empirical data on
trajectories of individuals in a flocking event, further compressed into signatures on the probability
simplex where mixed strategies of the game reside, these controls are found by solving an optimal
control problem to best-fit the signatures. The Pontryagin Maximum Principle (PMP) is used for this
computation. The time-averaged PMP Hamiltonian is our notion of cognitive cost of the event. In the
basic two mode version of the problem (on the one dimensional simplex) the formulation of the optimal
control problem makes the cognitive cost track the temporal variability of the signature and the control
– higher the variability higher is the cost.

We suggest that higher temporal variability of the event signature is an indicator of rapid morphological
changes to the flock as perceived by a predator, hence magnifying the confusion effect experienced by
the predator and leading to lowered success rate in capturing prey in the flock. This is supported by the
results of Section~\ref{sec:results}. We have also argued that another measure, the entropic cost, interpretable as how
uncertain the behavior of the flock appears to be to an observer (predator) has also some clues to offer,
despite the fact that it is insensitive to temporal variation. In the discussion of 2-dimensional simplex
signatures in Section~\ref{sec:energymodes}, the dominance of entropic cost of Event 2 draws attention to that case as
potentially influenced by predator attack. On the other hand data on the 1-dimensional simplex
signatures collected in Table~\ref{tbl:results} highlight the dominance of cognitive cost for Event 8, again a case
potentially influenced by predator attack. Taken together the cost measures suggest the possibility of
labeling Event 2 and Event 8 as predator attack related. 

In summary, the suggested measures of cost, especially the cognitive cost, may serve the purpose of extracting environmental influences from flock behavioral data. This is a tentative proposal and should be tested further, with optimal control studies with generative models of signature on higher dimensional simplex (beyond the 1-dimensional case studied in detail in Sections~\ref{sec:generative} and \ref{sec:results} of this paper).
One point to keep in mind is that the higher dimensional simplex offers a wealth of generative models
with multiple control modulation, but the time-averaged PMP Hamiltonian may still be the right notion
of cognitive cost. We aim to pursue these matters in future work. \\

\medskip
\noindent
\textbf{Data Accessibility.} Starling flight data was obtained from the COBBS group of the Institute for Complex
Systems in Rome (ISC-CNR), University of Rome “La Sapienza”, led by Andrea Cavagna. 

\smallskip
\noindent
\textbf{Authors’ Contributions.} U.H., V.R., and P.S.K. contributed to the theoretical conceptualization of the
cognitive cost and writing of the paper. M.M. and B.D. helped to edit the paper. M.M. and P.S.K. 
provided theoretical basis for energy mode decomposition. B.D. and P.S.K. developed techniques to generate
smooth trajectories from time-sampled raw data. U.H. and V.R. analyzed the smoothed starling flight
data. 

\smallskip
\noindent
\textbf{Competing Interests.} The authors declare no competing interests.

\bibliographystyle{IEEEtran}
\bibliography{reference}

\appendices
\renewcommand{\thefigure}{A-\arabic{figure}}
\setcounter{figure}{0}
\newpage
\section{Supplementary Material} \label{appdx:suppl}

\vspace*{20pt}
\begin{figure*}[!h]
\centering
	\includegraphics[ scale=0.35, trim = 0 0 0 25pt, clip = true]{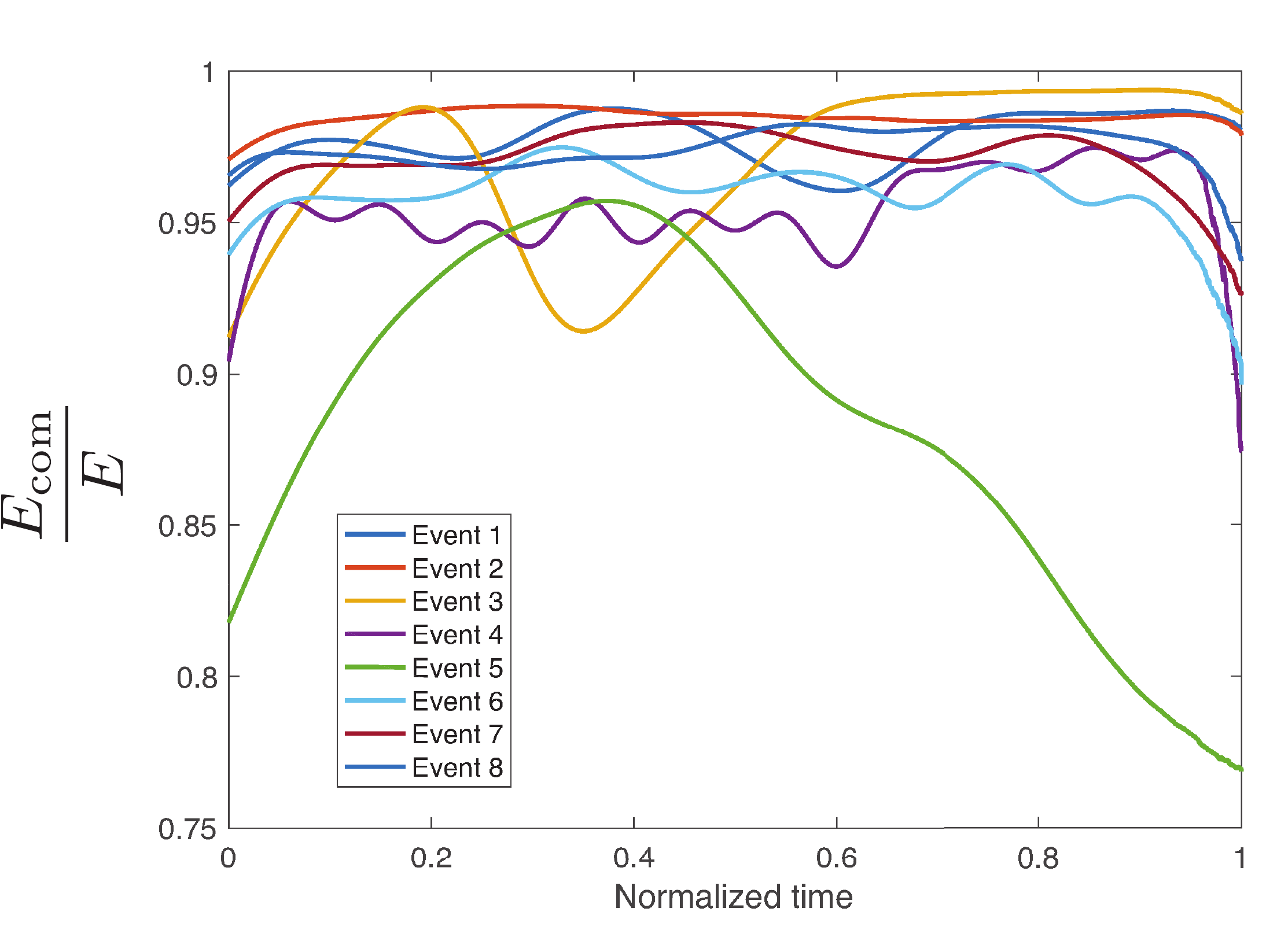}
	\caption{The fractions of the translational energy $E_{\text{com}}$ to the total kinetic energy $E$ for all eight events. These indicate $E_{\text{com}}$ is indeed the dominating part of the kinetic energy. }
	\label{fig:E_com}
	\vspace*{10pt}
\end{figure*}

\begin{figure*}[!ht]
\centering
	\includegraphics[width=\textwidth, trim = 0 15pt 0 0, clip = true]{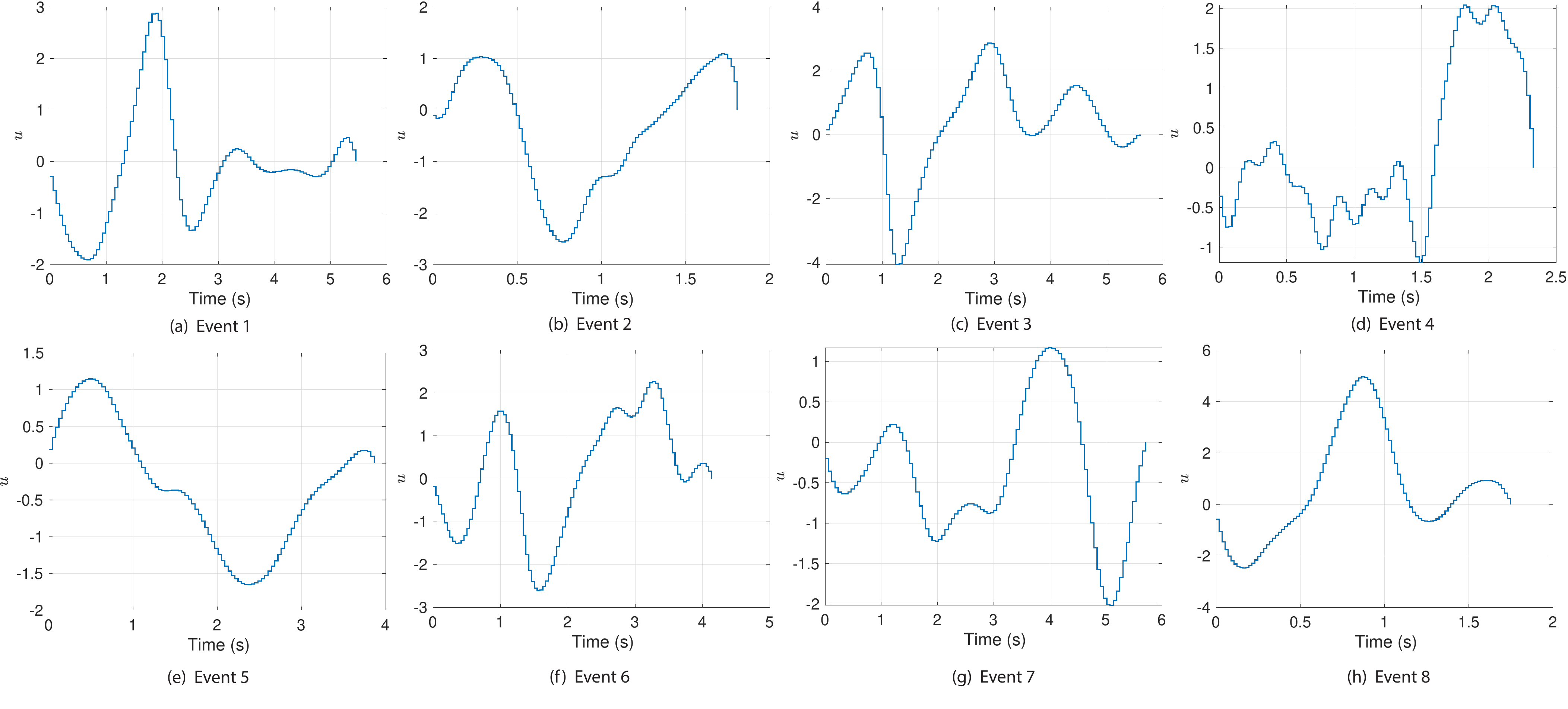}
	\caption{Optimal Controls to fit the energy signatures on the 1-D simplex for the case ENS-I}
	\label{fig:control_ens1}
	\vspace*{-10pt}
\end{figure*}
\begin{figure*}[h]
\centering
	\includegraphics[width=\textwidth, trim = 0 15pt 0 0, clip = true]{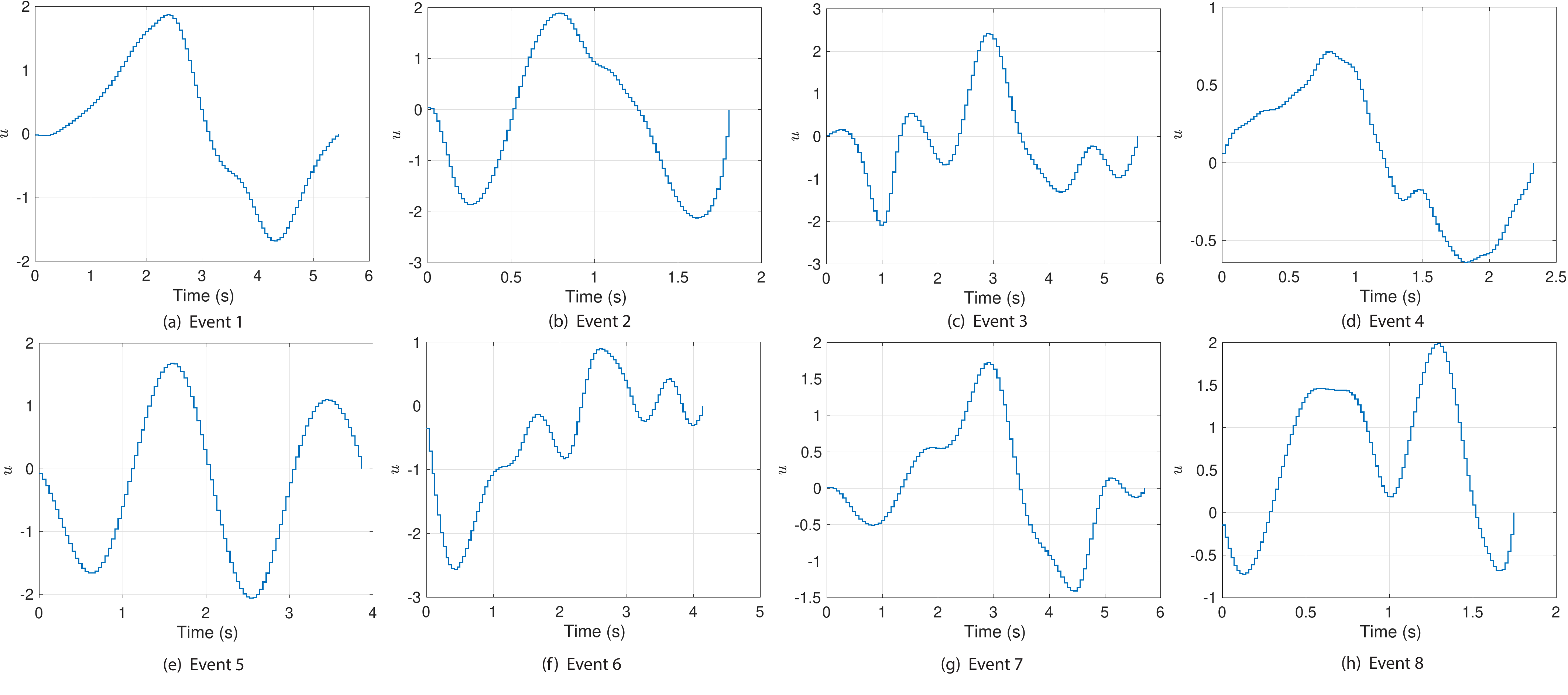}
	\caption{Optimal Controls to fit the energy signatures on the 1-D simplex for the case SHP-II}
	\label{fig:control_shp2}
\end{figure*}

\end{document}